\documentclass[final,5p]{elsarticle}

\usepackage{amssymb}
\usepackage{amsmath}
\usepackage{comment}
\usepackage{caption}
\usepackage{float}
\usepackage{subcaption}
\usepackage[T1]{fontenc}
\usepackage{graphicx}
\usepackage{siunitx}
\usepackage{amsmath}
\usepackage{booktabs}
\usepackage{url}

\usepackage{multirow}
\usepackage{dcolumn}
\usepackage{color}
\usepackage{lineno}
\usepackage{tikz}
\usepackage[inline]{enumitem}
\journal{Physics Letters B}

\begin{document}

\title{Constraining nucleon effective masses with flow and stopping observables from the S$\pi$RIT experiment}


\author[1,3,KENT]{C.Y.~Tsang\corref{cor1}}
\affiliation[1]{organization={National Superconducting Cyclotron Laboratory, Michigan State University, East Lansing, Michigan 48824, USA}}

\affiliation[3]{organization={Department of Physics, Michigan State University, East Lansing, Michigan 48824, USA}}
\affiliation[KENT]{organization={Kent State University, 800 E Summit St, Kent, Ohio, 44240, USA}}
\ead{ctsang1@kent.edu}
\cortext[cor1]{Corresponding author}
\author[2]{M.~Kurata-Nishimura\corref{cor1}}
\ead{mizuki@riken.jp}
\affiliation[2]{organization={RIKEN Nishina Center, Hirosawa 2-1, Wako, Saitama 351-0198, Japan}}

\author[1,3]{M.B.~Tsang\corref{cor1}}
\ead{tsang@frib.msu.edu}

\author[1,3]{W.G.~Lynch\corref{cor1}}
\ead{lynch@frib.msu.edu}

\author[4]{Y.X.~Zhang}
\affiliation[4]{organization={China Institute of Atomic Energy; Beijing, 102413, PR China}}

\author[1,3]{J.~Barney}

\author[1,3]{J.~Estee}

\author[1]{G.~Jhang} 


\author[1]{R.~Wang}

\author[2,5]{M.~Kaneko}
\affiliation[5]{organization={Department of Physics, Kyoto University, Kita-shirakawa, Kyoto 606-8502, Japan}}

\author[6]{J.W.~Lee}
\affiliation[6]{organization={Department of Physics, Korea University, Seoul 02841, Republic of Korea}}

\author[2]{T.~Isobe\corref{cor1}}
\ead{isobe@riken.jp}

\author[2,5]{T.~Murakami\corref{cor1}}
\ead{murakami.tetsuya.3e@kyoto-u.jp}

\author[2]{D.S.~Ahn}

\author[7,8]{L.~Atar}
\affiliation[7]{organization={Institut f\"ur Kernphysik, Technische Universit\"at Darmstadt, D-64289 Darmstadt, Germany}}
\affiliation[8]{organization={GSI Helmholtzzentrum f\"ur Schwerionenforschung, Planckstrasse 1, 64291 Darmstadt, Germany}}

\author[7,8]{T.~Aumann}

\author[2]{H.~Baba}

\author[8]{K.~Boretzky}

\author[10]{J.~Brzychczyk}
\affiliation[10]{organization={Faculty of Physics, Astronomy and Applied Computer Science, Jagiellonian University, Krak\'ow, Poland}}

\author[1]{G.~Cerizza}

\author[2]{N.~Chiga}

\author[2]{N.~Fukuda}

\author[2,9,7]{I.~Gasparic}
\affiliation[9]{Division of Experimental Physics, Rudjer Boskovic Institute, Zagreb, Croatia}

\author[6]{B.~Hong}

\author[7,8]{A.~Horvat}

\author[12]{K.~Ieki}
\affiliation[12]{organization={Department of Physics, Rikkyo University, Nishi-Ikebukuro 3-34-1, Tokyo 171-8501, Japan}}

\author[2]{N.~Inabe}

\author[13]{Y.J.~Kim}
\affiliation[13]{organization={Rare Isotope Science Project, Institute for Basic Science, Daejeon 34047, Republic of Korea}}

\author[14]{T.~Kobayashi}
\affiliation[14]{organization={Department of Physics, Tohoku University, Sendai 980-8578, Japan}}

\author[15]{Y.~Kondo}
\affiliation[15]{organization={Department of Physics, Tokyo Institute of Technology, Tokyo 152-8551, Japan}}

\author[11]{P.~Lasko}
\affiliation[11]{organization={Institute of Nuclear Physics PAN, ul. Radzikowskiego 152, 31-342 Krak\'ow, Poland}}

\author[13]{H.~S.~Lee}

\author[8]{Y.~Leifels}

\author[11]{J.~\L{}ukasik}

\author[1,3]{J.~Manfredi}

\author[16]{A.~B.~McIntosh}
\affiliation[16]{organization={Cyclotron Institute, Texas A\&M University, College Station, Texas 77843, USA}}

\author[1]{P.~Morfouace}

\author[15]{T.~Nakamura}

\author[2,5]{N.~Nakatsuka}

\author[2]{S.~Nishimura}

\author[2]{H.~Otsu}

\author[11]{P.~Paw\l{}owski}

\author[10]{K.~Pelczar}

\author[7,8]{D.~Rossi}

\author[2]{H.~Sakurai}

\author[1]{C.~Santamaria}

\author[2]{H.~Sato}

\author[7]{H.~Scheit}

\author[1]{R.~Shane}

\author[2]{Y.~Shimizu}

\author[8]{H.~Simon}

\author[17]{A.~Snoch}
\affiliation[17]{organization={Nikhef National Institute for Subatomic Physics, Amsterdam, Netherlands}}

\author[10]{A.~Sochocka}


\author[2]{T.~Sumikama}

\author[2]{H.~Suzuki}

\author[2]{D.~Suzuki}

\author[2]{H.~Takeda}

\author[1,KMUT,TACS]{S.~Tangwancharoen}
\affiliation[KMUT]{Department of Physics, Faculty of Science, King Mongkut's University of Technology Thonburi, Bangkok, Thailand}
\affiliation[TACS]{Center of Excellence in Theoretical and Computational Science (TACS-CoE), Faculty of Science, King Mongkut's University of Technology Thonburi, Bangkok, Thailand}

\author[7,8]{H.~T\"ornqvist}

\author[12]{Y.~Togano}

\author[18]{Z.~G.~Xiao}
\affiliation[18]{organization={Department of Physics, Tsinghua University, Beijing 100084, PR China}}

\author[16,19]{S.~J.~Yennello}
\affiliation[19]{organization={Department of Chemistry, Texas A\&M University, College Station, Texas 77843, USA}}

\author[18]{Y.~Zhang}

\author{S$\pi$RIT collaboration}





\date{\today}

\begin{abstract}
Properties of the nuclear equation of state (EoS) can be probed by measuring  the dynamical properties of nucleus-nucleus collisions. In this study, we present the directed flow ($v_1$), elliptic flow ($v_2$) and stopping (VarXZ) measured in fixed target Sn + Sn collisions at \SI{270}{AMeV} with the S$\pi$RIT Time Projection Chamber. 
We perform Bayesian analyses in which EoS parameters are varied simultaneously within the Improved Quantum Molecular Dynamics-Skyrme (ImQMD-Sky) transport code to obtain a multivariate correlated constraint. The varied parameters include symmetry energy, $S_0$, and slope of the symmetry energy, $L$, at saturation density, isoscalar effective mass, $m_{s}^*/m_{N}$, 
isovector effective mass, $m_{v}^{*}/m_{N}$ and the in-medium cross-section enhancement factor $\eta$. We find that the flow and VarXZ observables 
are sensitive to the splitting of proton and neutron effective masses and the in-medium cross-section. Comparisons of ImQMD-Sky predictions to the S$\pi$RIT data suggest a narrow range of preferred values for $m_{s}^*/m_{N}$, $m_{v}^{*}/m_{N}$ and $\eta$. 

\end{abstract}

\maketitle
\section{Introduction}

Nuclear matter is a significant component of neutron stars, and understanding its properties can elucidate many features of these celestial objects. Calculating the properties of both nuclear matter and neutron stars requires extensive knowledge of the nuclear equation of state (EoS), which describes the dependence of nuclear-matter internal energy on various state variables. Progress in understanding nuclear EoS has been achieved through heavy ion collisions~\cite{Xu2019, sorensen2023, LYNCH2022137098} and multimessenger astronomical observations of neutron stars~\cite{Danielewicz2002, LEFEVRE2016, Dutra2012, zhang2015electric, PREX2021, reed2021implications, brown2013, kortelainen2012nuclear, Danielewicz2016, tsang2009, MORFOUACE2019, Estee2021, cozma2018feasibility, RUSSOTTO2011471, Russotto2016}. In this paper, we present new experimental results on flow and stopping measurements from the S$\pi$RIT heavy ion collision experiment. Multiple observables are analyzed simultaneously using Bayesian inference to investigate correlations between various EoS parameters.

This paper is organized as follows: Section~\ref{sec:EOS} provides a brief overview of the nuclear EoS and relevant parameters. This is followed by a discussion of the experimental setup and the selection of observables in Section~\ref{sec:exp}. The transport model and Bayesian inference are discussed in Section~\ref{sec:Bay}. The experimental measurements and posterior constraints on EoS parameters are reported in Section~\ref{sec:result}, and finally, a summary is given in Section~\ref{sec:sum}.

\section{Nuclear equation of state}
\label{sec:EOS}

Nuclear EoS is a function of baryon number density $\rho$ and asymmetry $\delta = (\rho_n-\rho_p)/\rho$, where $\delta$ represents the difference in neutron ($\rho_n$) and proton ($\rho_p$)  number densities divided by total density $\rho$. We write the nuclear EoS as the sum of an isoscalar term $E_\text{is}(\rho)$ and an isovector term $E_\text{iv}(\rho, \delta)$, i.e. $E(\rho, \delta) = E_\text{is}(\rho, \delta)$ + $E_\text{iv}(\rho, \delta)$. The first term, $E_\text{is}(\rho)$, is the energy per nucleon of nuclear matter with equal proton and neutron densities ($\rho_p = \rho_n$); it provides the EoS of symmetric nuclear matter (SNM).  The second term describes how the energy changes as a function of neutron-proton asymmetry. It can be approximately written as $E_\text{iv}(\rho, \delta) = S(\rho)\delta^2 + \mathcal{O}(\delta^4)$, where $S(\rho)$ describes the dependence of nuclear EoS on neutron excess at different densities and is called the symmetry energy term. We truncate the expansion in $\delta$ at second order because the next (fourth) order term in $\delta$ contributes negligibly at asymmetries achieved in low energy nuclear collisions~\cite{Margueron2018}. 

Many current heavy-ion collision efforts have focused on constraining the first few coefficients in a Taylor expansion of $S(\rho)$ around saturation density, $\rho_0=\SI{0.16}{fm^{-3}}$. Such expansions are commonly parameterized by,
\begin{equation}
    S(\rho) = S_0 + Lx + \frac{1}{2}{K_\text{sym}}x^2 + \mathcal{O}(x^3),
    \label{eq:Taylor}
\end{equation}
where $x = (\rho - \rho_0)/3\rho_0$ and $S_0$,  $L$ and $K_\text{sym}$ are labels given to the first three expansion coefficients that describe the energy, slope and curvature of the EOS at saturation density, respectively.  Similarly, the isoscalar term is commonly parameterized as,
\begin{equation}
    E_\text{is}(\rho) = E_\text{0} + \frac{1}{2}K_\text{0}x^2 + \mathcal{O}(x^3),
    \label{eq:Taylor}
\end{equation}
where $E_\text{0}$ and $K_\text{0}$ are labels given to the first two non-zero expansion coefficients. From masses and other nuclide properties, the saturation energy for symmetric nuclear matter has been determined to be $E_\text{0} = -15.8\pm\SI{0.5}{MeV}$~\cite{Dutra2012}. Experiments that measured Giant Monopole resonances suggest that  $K_\text{0} = 230\pm\SI{30}{MeV}$~\cite{Dutra2012}.

Theoretical analysis has found that the form of momentum-dependent potential also affects $S(\rho)$~\cite{Zhang20202}. This momentum dependence can be quantified by ratios of the isoscalar effective mass, $m_s^*$, and isovector effective mass, $m_v^*$, to the mass of a nucleon, $m_N$,  in free space. The isoscalar effective mass comes from the isoscalar part of the momentum dependent mean field potential~\cite{Zhang20202}. In asymmetric matter, the strength of the neutron and proton effective mass splitting is related to the momentum dependence of the isovector mean-field potential.~\cite{Li2004, Brueckner1955, MAHAUX1985}. Near $\rho_{0}$, this splitting is related to the isovector effective mass $m_v^*/m_N = 1/(1 + \kappa)$, where $\kappa$ is the enhancement factor of the Thomas-Reiche-Kuhn sum rule~\cite{Zhang20202, Ring1980}.


The difference between the proton and neutron effective mass splitting, $\Delta m_{np}^*/\delta$, can be calculated from $m_v^*/m_N$ and $m_s^*/m_N$ with the following formula~\cite{li2018}, 

\begin{equation}
    \frac{\Delta m_{np}^*}{\delta} \approx -2 \left(\frac{m_N}{m_s^*} - \frac{m_N}{m_v^*}\right) \left(\frac{m_s^*}{m_N}\right)^2. 
    \label{eq:mvConvert}
\end{equation}

Recent measurements and analysis from the S$\pi$RIT experiment obtained a two-dimensional constraint on $\Delta m_{np}^*/\delta$ and $L$ through pion spectral ratio~\cite{Estee2021}. The yield ratio of $\pi^{-}$ to $\pi^{+}$ $p_T$ spectra is used to derive this constraint because both $\Delta m_{np}^*/\delta$ and $L$ influence the nucleon momenta; $L$ quantifies the isospin dependent contribution to the nucleon potential energy and $\Delta m_{np}^*$ quantifies the isospin dependent impact on the nucleon kinetic energy. Either increasing $L$ or decreasing  $\Delta m_{np}^*$ will increase the energies of neutrons relative to protons. This increases the numbers of n-n collisions relative to p-p collisions at energies above the pion production threshold and enhances the production of $\pi^{-}$ relative to that of $\pi^{+}$.


\section{Experimental setup and observable selection}
\label{sec:exp}

\subsection{Experimental setup}
In the S$\pi$RIT experiment, we bombarded isotopically enriched $^{112}$Sn and $^{124}$Sn targets with secondary radioactive $^{108}$Sn and $^{132}$Sn beams and also stable $^{112}$Sn and $^{124}$S beams at \SI{270}{AMeV}. The targets were  placed at the entrance of the S$\pi$RIT Time Projection Chamber (TPC), which was installed inside the SAMURAI dipole magnet~\cite{Barney2019, Shane2015} at the Radioactive Isotope Beam Factory (RIBF). The S$\pi$RIT TPC identified and measured the  momenta of charged particles~\cite{estee2019, Shane2015, LEE2020, Barney2019} produced in $^{108}$Sn$ + ^{112}$Sn, $^{112}$Sn$ + ^{124}$Sn, $^{124}$Sn$ + ^{112}$Sn and $^{132}$Sn +$^{124}$Sn collisions. Some results for the production of  pions, hydrogen and helium isotopes have been previously published ~\cite{JHANG2021,Estee2021, Masanori2022, Lee2022}. In this paper, we present analyses of collective flow and stopping from this experiment.

\subsection{Observable selection}
\label{sec:obsSec}
Collective flow is a descriptive label for a group of observables that have been widely used to constrain the nuclear EoS using heavy ion collisions~\cite{Danielewicz2002, RAMI1999, Andronic2001, Filippo2017,le2016constraining, Russotto2016,Russotto2023}. It often involves analyses of anisotropies in the azimuthal distributions of emitted particles with respect to the reaction plane. Such collective flow observables in nucleus-nucleus collisions commonly reflect the pressures on \emph{participant nucleons} in the overlapping region of projectile and target wherein this participant matter is compressed. Flow observables also reflect the presence of \emph{spectator nucleons}  that reside outside of the participant region and block the escape of participant nucleons from the compressed participant region. 

Flow is a promising observable to constrain nuclear EoS because of its correlation with nuclear pressure. If the mean field is highly repulsive, participant nucleons experience higher pressures which leads to early emission, but this emission is partially blocked by the spectator nucleons if they have not already moved past the participant region before it can expand into the spectator matter~\cite{Danielewicz2002, Stoicea2004, Reisdorf2012}. The blocking of the expanding participant matter by the spectator nucleons results in azimuthal anisotropies in fragment emissions. In very central collisions, there is very little spectator matter so emitted particles exhibit little anisotropies. With increasing impact parameter, the amount of spectator matter increases and the importance of the spectators blocking the emitted particles results in the increasing directional dependence that is characteristic of the directed flow. 

Collective flow can be quantified by the Fourier coefficients of the fragments' azimuthal distributions with respect to the azimuthal angle for the reaction plane $\Phi$~\cite{Voloshin1996},
\begin{equation}
    \frac{dN}{d(\phi - \Phi)} \propto 1 + 2v_1\cos(\phi - \Phi) + 2v_2\cos(2(\phi - \Phi)) + ...
\end{equation}
In the above equation, $N$ is the particle yield, $\phi$ is the azimuthal angle of emission for the particle, $v_1$ is called the \emph{directed flow} and $v_2$ is called the \emph{elliptic flow}. Experimentally, $v_1$ and $v_2$ are calculated by the following formula,

\begin{equation}
    \begin{split}
        v_1 &= \langle \cos(\phi - \Phi)\rangle, \\
        v_2 &= \langle \cos(2(\phi - \Phi))\rangle. \\
    \end{split}
    \label{eq:vobs}
\end{equation}

In this paper, we determined the azimuthal angle $\Phi$ of the reaction plane experimentally with the Q-vector method~\cite{Poskanzer1998}. Q-vector is defined as,

\begin{equation}
    \vec{Q} = \sum_{i=0}^{N} w_i\hat{p_T}_i\ \text{sign}(y_{0i}),
    \label{eq:Qvec}
\end{equation}
where $\hat{p_T}_i$ is a unit vector pointing in the direction of the transverse momentum of the $i^\text{th}$ track, $y_{0i} = (y_{CM}/y_\text{NN})_i$ is the $i^\text{th}$ particle's rapidity in the C.M. frame ($y_\text{CM}$) normalized by beam rapidity in the nucleon-nucleon frame ($y_\text{NN} = 0.5y_\text{beamLab}$), and sign$(x)$ is the sign function. We are free to choose the weighting factor $w_i$, with common choices including $w_i = 1$ or $w_i = p_T$. The effect of using different $w_i$ will be considered as systematic uncertainty. The reaction plane angle $\Phi$ is chosen by the azimuthal angle of $\vec{Q}$. Although this approximation and the limited detector acceptance causes non-negligible broadening in the reaction plane resolution, appropriate formulas are used to correct for these effects. For details on these corrections, please refer to Ref.~\cite{Poskanzer1998}. In this manuscript, we report only the flow values that have been corrected.

Equation~\eqref{eq:vobs} can be calculated by averaging over fragments of the same species. Both the theoretical and experimental values of $v_1$ and $v_2$ depend on the mass and species of each fragment, but the probability of producing a cluster of a particular mass depends on the details of the clusterization mechanism of each transport model. The underlying physics of this process is not accurately calculated by most transport codes~\cite{ONO2019, Donigus2020}, and this can result in significant systematic uncertainties in theoretical predictions of flow for different isotopes.

To compute light fragments, such as deuterons, tritons, and helium from final nucleon distribution from transport models, various cluster recognition methods have been employed, such as the minimum spanning tree method used in QMD type models~\cite{Gossiaux1997}. This method classifies neutrons and protons that are emitted at small relative distances and momenta as heavy clusters.~\cite{Gossiaux1997}. However, this process has a model dependence that reflects the influence of long-range multi-particle correlations that are not yet fully understood~\cite{ONO2019, Donigus2020}. As a result, isotope-specific observable heavily depends on a detailed understanding of clusterization, and transport model predictions for these observables can often be unreliable.

To construct an observable that does not require an accurate description of the clusterization process,
we calculate the \emph{Coalescence Invariant flow} (C.I. flow) distributions. These distributions approximate the flow of nucleons prior to cluster formation by including contributions from p, d, t,$^3$He and $^4$He together in the calculation of averages of cosines in Eq.~\eqref{eq:vobs}. Each fragment is weighted by their number of protons, i.e. Helium isotopes are weighted twice as much as Hydrogen isotopes. Fragments heavier than $^4$He are not included due to their low yields. 

We select the impact parameter with gates on total detected charged particle multiplicity as described in Ref.~\cite{Barney2019}. This centrality selection method was also used in our previous S$\pi$RIT publications~\cite{JHANG2021,Estee2021,Masanori2022}. Due to the limited geometric acceptance in the S$\pi$RIT TPC, nuclear fragments with large momenta emitted at backward angles in the C.M. frame cannot be efficiently detected, so we limit our flow data to $0 \leq y_0 \leq 0.8$. 

In addition to the mean field potentials in the EoS, momentum transfers that contribute to collective flow are also influenced by the in-medium nucleon-nucleon (NN) cross-section~\cite{Chen20212, LI2022}.  We construct the stopping observable, \emph{VarXZ}=\emph{VarX}/\emph{VarZ}, where \emph{VarX} and \emph{VarZ} are the variances of particle rapidity distributions in the transverse and longitudinal directions, respectively.   Since VarXZ is a ratio of variances, much of the systematic error from clusterization is cancelled out in the division. VarXZ measures the degree of stopping and thermalization~\cite{REISDORF2010}, and has also been used to probe the nuclear shear viscosity~\cite{GAITANOS2005}. It is closely influenced by the in-medium cross-section~\cite{REISDORF2010}.  

To reconstruct this observable and calculate the momentum of the particles accurately, only tracks emitted nearly perpendicular to the magnetic field of the TPC are used. Based on the performance of the S$\pi$RIT TPC~\cite{Barney2021}, azimuth cuts of $330^{\circ} < \phi < 360^{\circ}$, $0^{\circ} < \phi < 20^{\circ}$ and $160^{\circ} < \phi < 210^{\circ}$ are used for this purpose. These cuts are also used in the other S$\pi$RIT analyses~\cite{JHANG2021, Estee2021, Kaneko2021, Lee2022}. Since determination of the reaction plane does not require as precise values for the magnitude of the momenta of particles, and to minimize bias due to particle cut, we do not impose these restrictive cuts in azimuthal angle on particles in calculating azimuthal orientation of the reaction plane. 

The $x$-axis in \emph{VarX} can be any arbitrary laboratory axis that is perpendicular to the beam axis, consistent with definitions in Ref.~\cite{REISDORF2010}. Given the arbitrary azimuthal orientation of the $x$-axis for the \emph{VarX} observable, we can and do calculate $x$-rapidity distribution by projecting $p_T$ of each track onto planes with random azimuthal angles. 

S$\pi$RIT TPC cannot efficiently measure fragments at $y_0 < 0$ so data is not available at all rapidities for $^{108}$Sn$ + ^{112}$Sn and $^{132}$Sn$ + ^{124}$Sn reactions, individually. However, we have constructed the full rapidity distribution of $^{112}$Sn$ + ^{124}$Sn by combining the results of $^{112}$Sn$ + ^{124}$Sn and $^{124}$Sn$ + ^{112}$Sn reactions. Our evaluation of VarXZ is limited only to this reaction system as the other systems do not have the corresponding mirror reactions. In the following, we select central events of $\langle b\rangle = \SI{1}{fm}$ in order to maximize contributions from nucleon-nucleon collisions. The absence of the flow from $^{112}$Sn$ + ^{124}$Sn should impact our conclusion minimally since the $\delta$ value of $^{112}$Sn$ + ^{124}$Sn is between that of $^{108}$Sn$ + ^{112}$Sn and $^{132}$Sn$ + ^{124}$Sn systems, therefore the range of asymmetry being studied is not affected.

\begin{table}[h]
\begin{center}
\caption{List of observables used in this analysis. The second column (Exp. $\langle b\rangle$) shows the averaged impact parameters of the selected experimental events in reconstruction of the observable. }
\label{tab:expList}
\begin{tabular*}{0.8\linewidth}{@{\extracolsep{\fill}}lcc}
\toprule
Observable                                                 & Exp. $\langle b\rangle$ &  System                  \\
\hline
\\[-1ex]
\multirow{2}{*}{C.I. $v_1$ v.s. $y_0$}                     & 5.0 fm                    & $^{108}$Sn$ + ^{112}$Sn \\
                                                           & 5.0 fm                    & $^{132}$Sn$ + ^{124}$Sn \\
\\[-1ex]    
C.I. $v_1$ v.s. $p_T$                                       & 5.0 fm                     & $^{108}$Sn$ + ^{112}$Sn \\
    ($0.3 < y_0 < 0.8$)                                     & 5.0 fm                    & $^{132}$Sn$ + ^{124}$Sn \\
\\[-1ex]
\multirow{2}{*}{C.I. $v_2$ v.s. $y_0$}                     & 5.0 fm                    & $^{108}$Sn$ + ^{112}$Sn \\
                                                           & 5.0 fm                     & $^{132}$Sn$ + ^{124}$Sn \\
\\[-1ex]                                                           
VarXZ                                                      & 1.0 fm                    & $^{112}$Sn$ + ^{124}$Sn \\
\bottomrule
\end{tabular*}
\end{center}
\end{table}

Table~\ref{tab:expList} summarizes all observables for the corresponding reactions that will be used in this manuscript for comparison with models. Fragments at mid-rapidity have small $v_1$, so the rapidity range of $v_1$ when being plotted as a function of $p_T$ is narrowed down to $0.3 < y_0 < 0.8$ to enhance the sensitivity. Their average impact parameters selected from multiplicity gates are shown in the second column and the corresponding reaction systems are shown in the third.

\subsection{Observable uncertainties}

Three independent sources of systematic uncertainties are considered in the reconstruction of C.I. flow: 1) the variation of the Q-vector weighting conditions, 2) the variation of track selection and 3) variation of impact parameter selection. To quantify the uncertainty due to the variation in the Q-vector weights $w_i$, we reconstruct each flow spectrum multiple times using different forms of $w_i$ for Q-vector calculation. Next, we fix the weights in the Q-vector and vary the goodness of track selection conditions to estimate the systematic uncertainty from the second source, which includes varying the minimal track length threshold and azimuth cuts. Finally, we reconstruct the observable with a different multiplicity gate. The upper and lower limits of the multiplicity gate is varied in such a way that the average impact parameter remains unchanged. We determine the systematic uncertainty as the bin-by-bin difference between spectra constructed using the default and varied multiplicity gate.

For VarXZ, systematic uncertainty is considered by varying track and impact parameter selection. The total observable uncertainty for both VarXZ and flow is obtained  by adding the statistical uncertainty and all sources of systematic uncertainties in quadrature. The experimental values and uncertainties are presented in~\ref{app:data}.

\section{Transport model and Bayesian inference}
\label{sec:Bay}

All measurements in Table~\ref{tab:expList} are compared simultaneously to predictions from the Improved Quantum Molecular Dynamic-Skyrme model (ImQMD-Sky)~\cite{ZHANG2014,ZHANG2015}, which has been frequently used to study nucleus-nucleus collisions at similar beam energies~\cite{MORFOUACE2019}. In the model, the nucleonic potential energy density without the spin-orbit term is given by the sum of two terms, $u_\text{loc}+u_\text{md}$, where,
\begin{equation}
    \begin{split}
\label{eq:edfimqmd}
u_\text{loc}=&\frac{\alpha}{2}\frac{\rho^2}{\rho_0} +\frac{\beta}{\gamma+1}\frac{\rho^{\gamma+1}}{\rho_0^\gamma}+\frac{g_\text{sur}}{2\rho_0 }(\nabla \rho)^2+\\
&\frac{g_\text{sur,iso}}{\rho_0}[\nabla(\rho_n-\rho_p)]^2+A_\text{sym}\frac{\rho^2}{\rho_0}\delta^2+B_\text{sym}\frac{\rho^{\gamma+1}}{\rho_0^\gamma}\delta^2.
    \end{split}
\end{equation}
Parameters will be defined in the next paragraph. The Skyrme-type momentum-dependent energy density functional, $u_\text{md}$ stems from an interaction of the form $\delta (\mathbf r_1-\mathbf r_2 ) (\mathbf p_1-\mathbf p_2 )^2$~\cite{Skyrme1956,Vautherin1972,ZHANG2015} and is written as,

\begin{equation}
    \begin{split}
\label{eq:mdimqmd}
u_\text{md}=&C_0\sum_{ij}\int d^3pd^3p' f_i(\mathbf r,\mathbf p)f_j(\mathbf r,\mathbf p')(\mathbf p-\mathbf p')^2+\\
&D_0\sum_{ij\in \text{n}}\int d^3pd^3p'f_i(\mathbf r,\mathbf p) f_j(\mathbf r,\mathbf p')(\mathbf p-\mathbf p')^2 +\\
&D_0\sum_{ij\in \text{p}}\int d^3p d^3p' f_i(\mathbf r,\mathbf p)f_j(\mathbf r,\mathbf p')(\mathbf p-\mathbf p')^2.
    \end{split}
\end{equation}
There are nine parameters in Equations~\eqref{eq:edfimqmd} and~\eqref{eq:mdimqmd}, which are $\alpha$, $\beta$, $\gamma$, $A_\text{sym}$, $B_\text{sym}$, $C_0$, $D_0$, $g_\text{sur}$ and $g_\text{sur;iso}$. Calculations show that the predicted observables are relatively insensitive to $g_\text{sur}$ and $g_\text{sur,iso}$ ~\cite{Zhang20202}. We therefore set them to $g_\text{sur} = \SI{24.5}{MeV\ fm^2}$ and $g_\text{sur,iso} = \SI{-4.99}{MeV\ fm^2}$. These parameters are the same as those derived from the commonly used Skyrme interaction  SLy4~\cite{Chabanat1997}. 

The remaining seven free parameters are related to the seven nuclear EoS parameters ($\rho_0$, $E_\text{sat}$, $K_\text{sat}$, $S_0$, $L$, $m_s^*$ and $ m_v^*$) through appropriate formulas from Refs.~\cite{Agrawal2005,Chen2009}. The saturation density and coefficients of isoscalar terms are well constrained from previous studies, so they are fixed to $\rho_0=\SI{0.155}{fm^{-3}}$, $E_\text{0} = \SI{-15.8}{MeV}$ and $K_\text{0} = \SI{230}{MeV}$~\cite{Dutra2012}. The four remaining EoS parameters ($S_0$, $L$, $m_s^*/m_N$ and $m_v^*/m_N$) and the in-medium cross-section enhancement factor $\eta$ (defined in the next paragraph) are varied in this study. 

In ImQMD-Sky, the in-medium NN cross section is formulated in a phenomenological form,

\begin{equation}
    \sigma^\text{med}_\text{QMD} = \left (1 - \frac{\eta\rho}{\rho_0}\right )\sigma^\text{free},
\label{eq:imcs}
\end{equation}
where $\sigma^\text{free}$ is the NN cross-section in free space from Ref.~\cite{Cugnon1996} and $\eta$ is the enhancement factor to be determined. Note that with this definition, $\eta$ has the opposite sign than what was used previously in Ref.~\cite{Chen20212}. In Eq.~\eqref{eq:imcs},  positive $\eta$ implies that the in-medium cross section is reduced from that for free nucleon-nucleon scattering.

\emph{Bayesian inference} is performed to convert experimental results into correlated constraints on all five varying parameters. This analysis requires \emph{prior} distributions as input, which encodes our initial belief in parameter values from previous studies. The analysis returns a multivariate distribution called the \emph{posterior}, which is the probability distribution of parameter values conditioned on the measured observables using the Bayes theorem. Let $\vec{x}$ be the list of parameters and $\vec{O}$ be the list of measured observable values. Bayes theorem then states that,

\begin{equation}
    P(\vec{x}|\vec{O}) \propto P(\vec{x})P(\vec{O}|\vec{x}).
    \label{eq:Bayes}
\end{equation}

In this equation, $P(\vec{x}|\vec{O})$ is the posterior, $P(\vec{x})$ is the prior and $P(\vec{O}|\vec{x})$ is called the \emph{likelihood}, which is the probability of getting the observed results provided that parameter values in set $\vec{x}$ are the true values. Likelihood is usually modelled as,

\begin{equation}
\begin{split}
    &P(\vec{O}|\vec{x}) \propto \\
    &\exp{\left(-(\vec{O} - \vec{O}^\text{model}(\vec{x}))\Sigma^{-1}(\vec{O} - \vec{O}^\text{model}(\vec{x}))^T\right)},
\end{split}
\end{equation}
where $\vec{O}$ is the list of all measured observables arranged as a vector, $\vec{O}^\text{model}(\vec{x})$ is the predicted values from theoretical model for a given parameter set $\vec{x}$ and $\Sigma = \Sigma(\vec{O}^\text{model}) + \text{diag}(\sigma_O)$ is the combined covariance matrix for theoretical and experimental uncertainties, with the first term $\Sigma(\vec{O}^\text{model})$ denoting the covariance matrix of all model predictions and $\text{diag}(\sigma_O)$ is a diagonal matrix with experimental uncertainty as the diagonal values.

In this work, Markov chain Monte Carlo (MCMC) from the PyMC2 package~\cite{Patil2010} is used to compute the posterior distribution. To speed up the MCMC process, we employ Gaussian emulators~\cite{Rasmussen2005} and Principal Component Analysis to efficiently interpolate predictions from ImQMD-Sky on just 70 parameter sets and estimate model covariance $\Sigma(\vec{O}^\text{model})$. Two Principal Components are used for each observable, capturing more than 95\% of the variance in all the training spectrum. These parameter sets are sampled uniformly and randomly on a Latin hypercube within the parameter ranges given in Table~\ref{tab:parRange}. The ranges of $L$, $m_s^*/m_N$ and $m_v^*/m_N$ are maximal, meaning that beyond these ranges ImQMD-Sky is unable to simulate correctly. For each parameter set and system in Table~\ref{tab:expList}, the training spectrum are generated from 20,000 ImQMD-Sky simulated collisions. The simulation time of ImQMD-Sky is \SI{400}{fm/c}, long after the time when the collision dynamics finishes and the observables reach their asymptotic values. We have verified that and our observables reach an asymptotic values when the simulation time is greater than \SI{300}{fm/c}. The final MCMC posterior is generated with 630,000 steps.

\begin{table}[h]
\begin{center}
\caption{The ranges of parameters for Bayesian analysis. The last two columns show the Gaussian prior mean and standard deviation ($\sigma$) for $S_0$ and $L$. The last three parameters have uniform prior so no numbers are provided. }
\label{tab:parRange}
\begin{tabular*}{0.8\linewidth}{@{\extracolsep{\fill}}lcccc}
\toprule
Parameters & Min. & Max. & Mean & $\sigma$\\
\hline
\\[-1ex]
$S_0$ (MeV)      & 25   & 52   & 35.3 & 2.8\\
$L$ (MeV)        & 18   & 160  & 80 & 38\\
$m_s^*/m_N$      & 0.6  & 1    & \multicolumn{2}{c}{Uniform}\\
$m_v^*/m_N$      & 0.6  & 1.2  & \multicolumn{2}{c}{Uniform}\\
$\eta$           & -0.3 & 0.3  & \multicolumn{2}{c}{Uniform}\\
\bottomrule
\end{tabular*}
\end{center}
\end{table}

\begin{figure*}[!ht]
    \centering
    \includegraphics[width=0.9\linewidth, trim={0 5cm 0 2.5cm},clip]{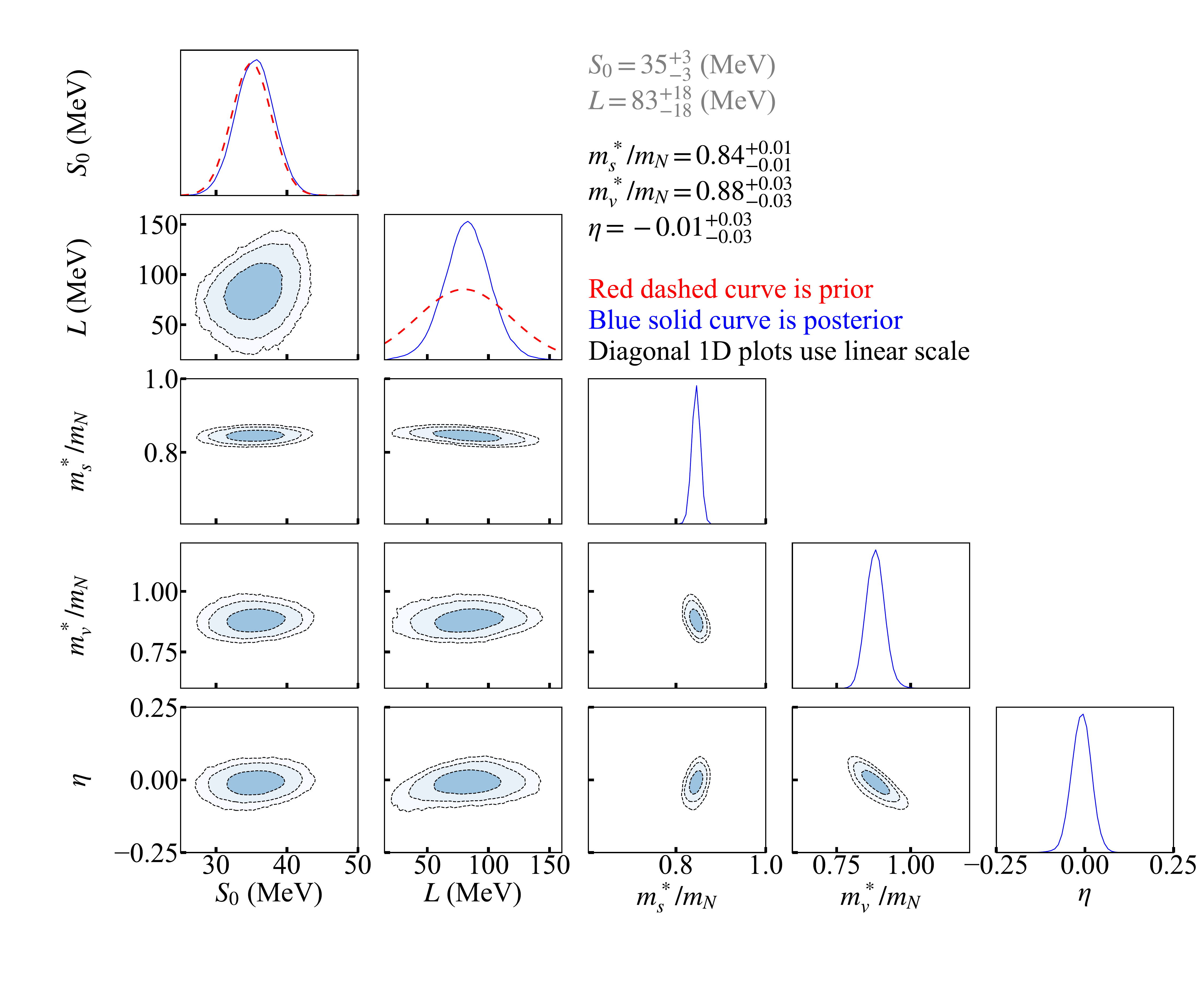}
    \caption{Posterior distribution after ImQMD-Sky's predictions are compared to experimental data with Bayesian analysis using MCMC. The median and 68\% confidence interval of the parameters are listed on the upper right hand side of the figure. The numbers for $S_0$ and $L$ are grayed out to signify their dependence on prior. See text for details. Priors of $S_0$ and $L$ are plotted on the diagonal plot as red dashed curves, and the priors of the other parameters are all uniform so they are not plotted for simplicity. The outer boundaries for the three shaded blue regions in the off-diagonal plots, from the deepest shade to the lightest of blue, correspond to the 68\%, 95\% and 99\% confidence intervals. }
    \label{fig:ExpPosterior}
\end{figure*}

When five parameter sets are removed from interpolation, the interpolation error does not increase, which indicates that 70 sets are more than sufficient. Gaussian priors of $S_0\sim\text{Gaus}(\mu=35.3, \sigma=2.8)$ MeV and $L\sim\text{Gaus}(\mu=80, \sigma=38)$ MeV are used while uniform priors within the experimental known ranges are used for the effective masses and $\eta$. The priors on $S_0$ and $L$ come from the posterior distributions from the analysis of pion spectral ratios in the same experiment~\cite{Estee2021}.

ImQMD-Sky calculations are done at $b = \SI{5}{fm}$ for flow observables and $b = \SI{1}{fm}$ for stopping observables. They are chosen to match the mean impact parameters of the observables in Table~\ref{tab:expList}. Note that the range of possible impact parameters that contribute to the experimental measurements actually varies due to multiplicity fluctuations. Ideally, we should simulate events with a realistic distribution of impact parameters and apply the same multiplicity cut as data. Unfortunately, the multiplicity distributions of most transport models  are not precisely comparable to data due to issues with coalescence algorithms described in Section~\ref{sec:obsSec}.

\section{Analysis results}
\label{sec:result}

The posterior is shown in Fig.~\ref{fig:ExpPosterior}. Despite employing uniform priors, tight constraints on $m_s^*/m_N$, $m_v^*/m_N$ and $\eta$ are observed. These constraints are robust constraints as they do not show a significant correlation with either $S_0$ or $L$, indicating that their posterior values remain unaffected by our choice of prior.

In contrast, the posterior distributions of $S_0$ and $L$ are wide even when Gaussian priors are used. The posterior distribution of $S_0$ looks very similar to the prior, indicating that our choice of observables lacks sensitivity to $S_0$. Although there is a modest improvement in the constraint on $L$ compared to the prior distribution, Fig.~\ref{fig:ExpPosterior} also highlights a correlation between $S_0$ and $L$. Consequently, the marginalized value of $L$ ($83\pm\SI{18}{MeV}$) may change if a different prior for $S_0$ is employed.

The near-zero value of $\eta = -0.01^{+0.03}_{-0.03}$ indicates that the in-medium cross-section is similar to the free cross-section. This differs from previous analyses performed at different beam energies, where an enhancement in the in-medium cross-section is derived ~\cite{Chen20212}. 

\begin{figure}[!ht]
    \centering
    \begin{subfigure}[b]{0.45\linewidth}
        \centering
        \includegraphics[width=\linewidth]{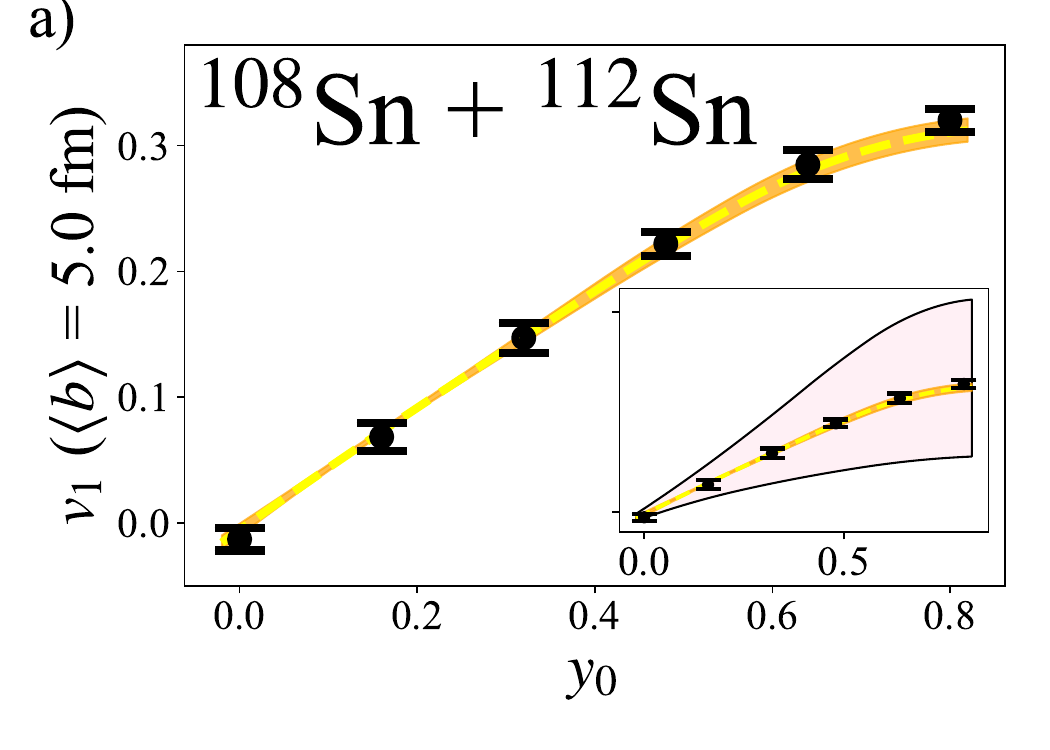}
    \end{subfigure}%
    \begin{subfigure}[b]{0.45\linewidth}
        \centering
        \includegraphics[width=\linewidth]{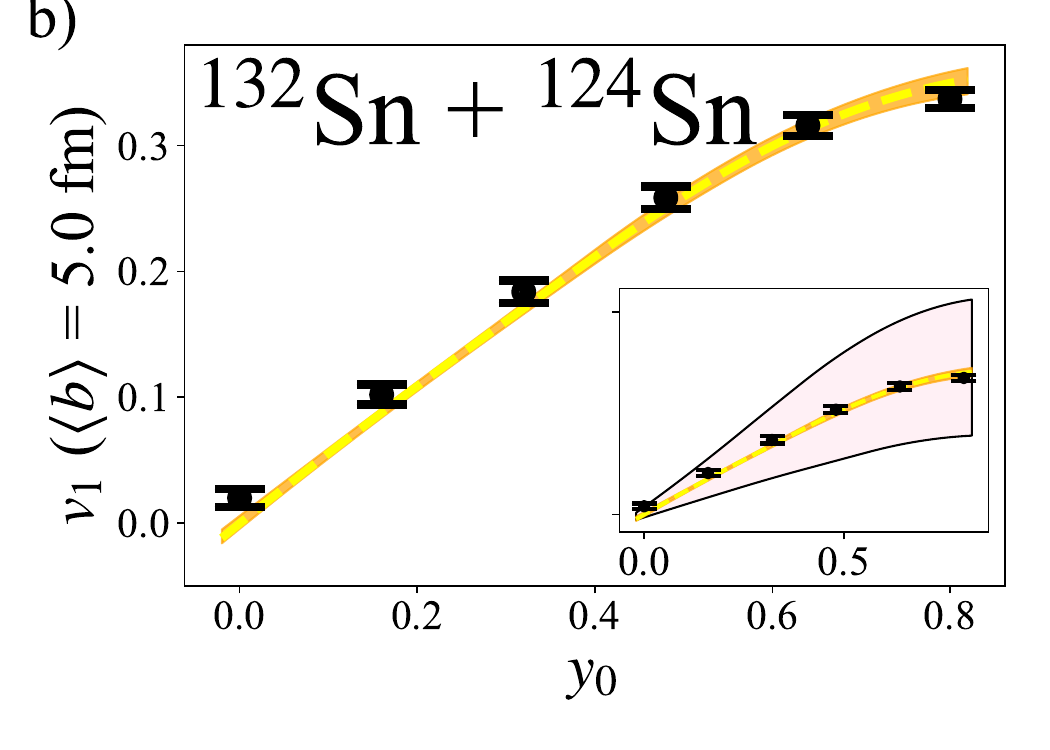}
    \end{subfigure}
        \begin{subfigure}[b]{0.45\linewidth}
        \centering
        \includegraphics[width=\linewidth]{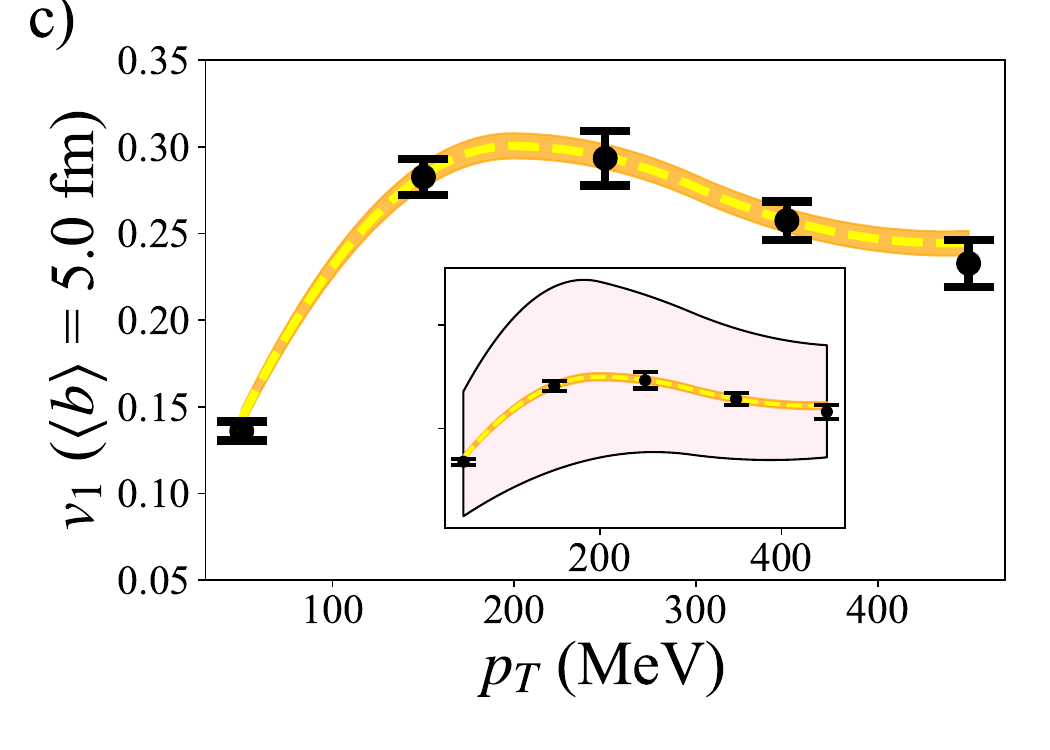}
    \end{subfigure}%
    \begin{subfigure}[b]{0.45\linewidth}
        \centering
        \includegraphics[width=\linewidth]{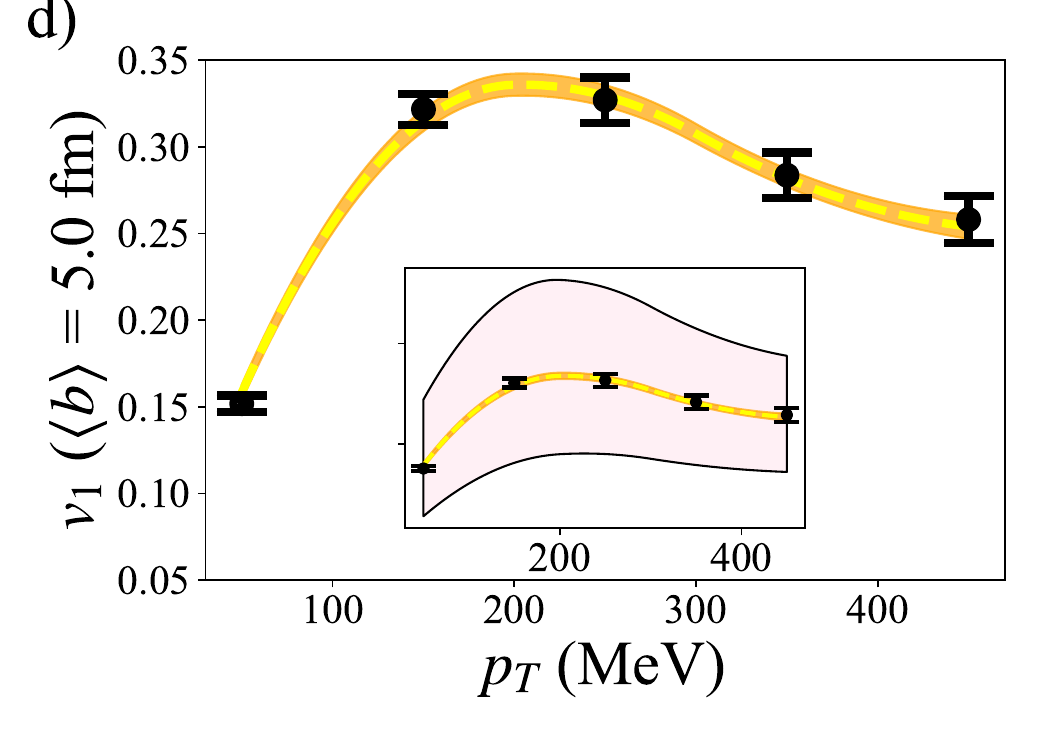}
    \end{subfigure}
    \begin{subfigure}[b]{0.45\linewidth}
        \centering
        \includegraphics[width=\linewidth]{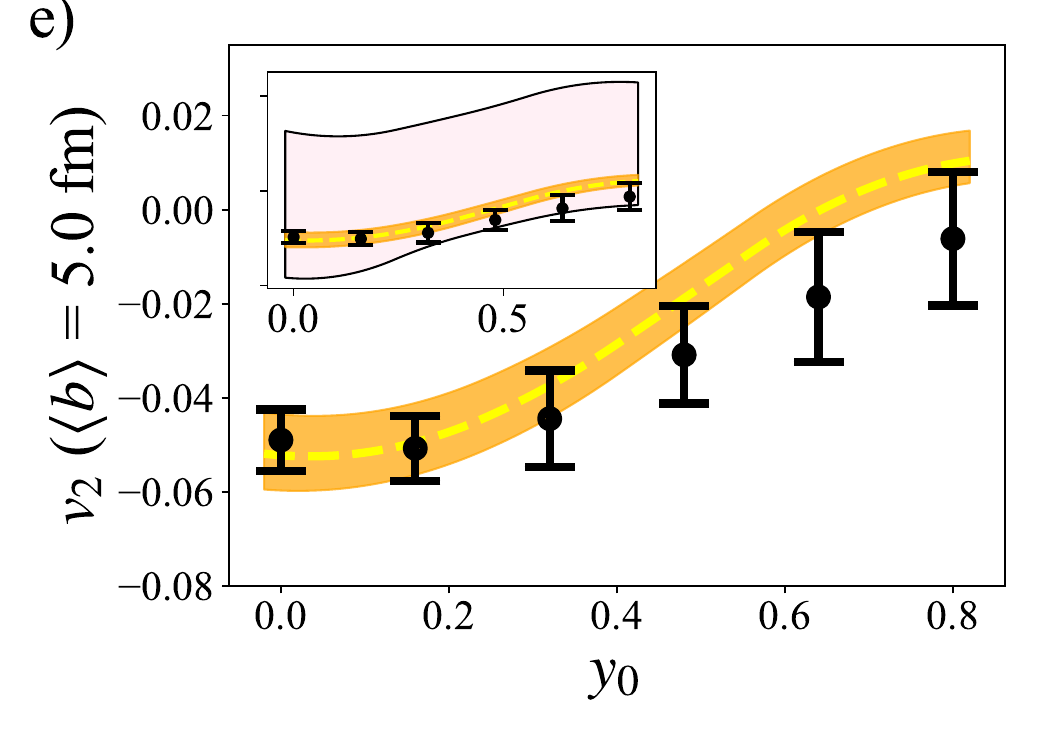}
    \end{subfigure}%
    \begin{subfigure}[b]{0.45\linewidth}
        \centering
        \includegraphics[width=\linewidth]{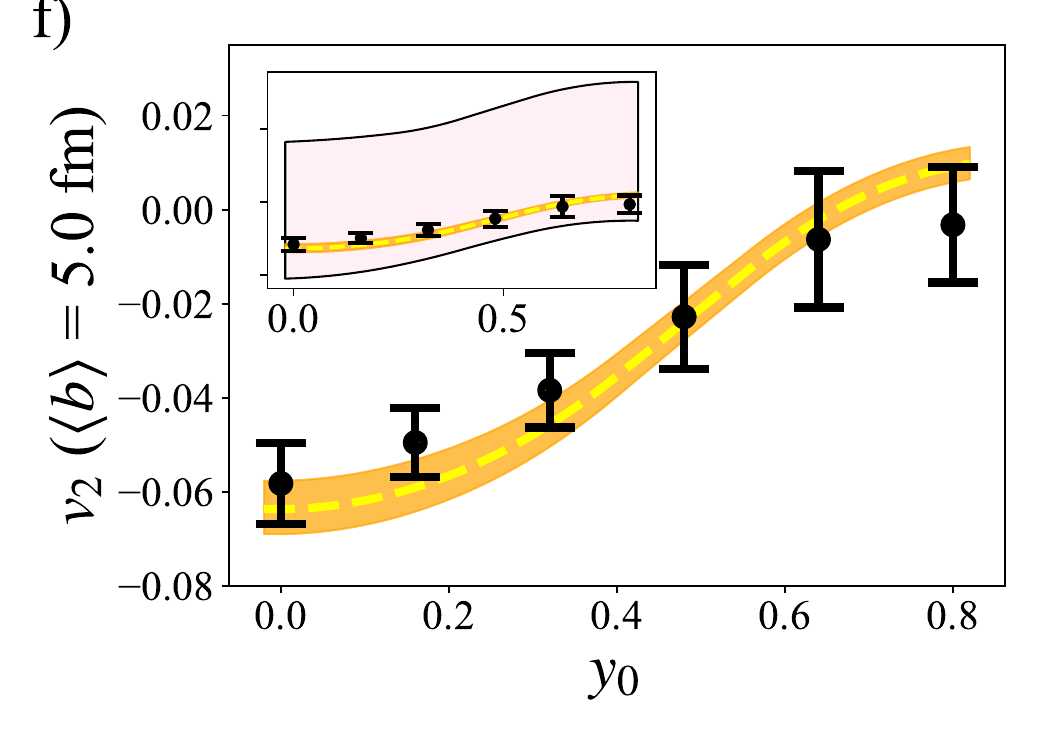}
    \end{subfigure}
    \begin{subfigure}[b]{0.65\linewidth}
        \centering
        \includegraphics[width=\linewidth]{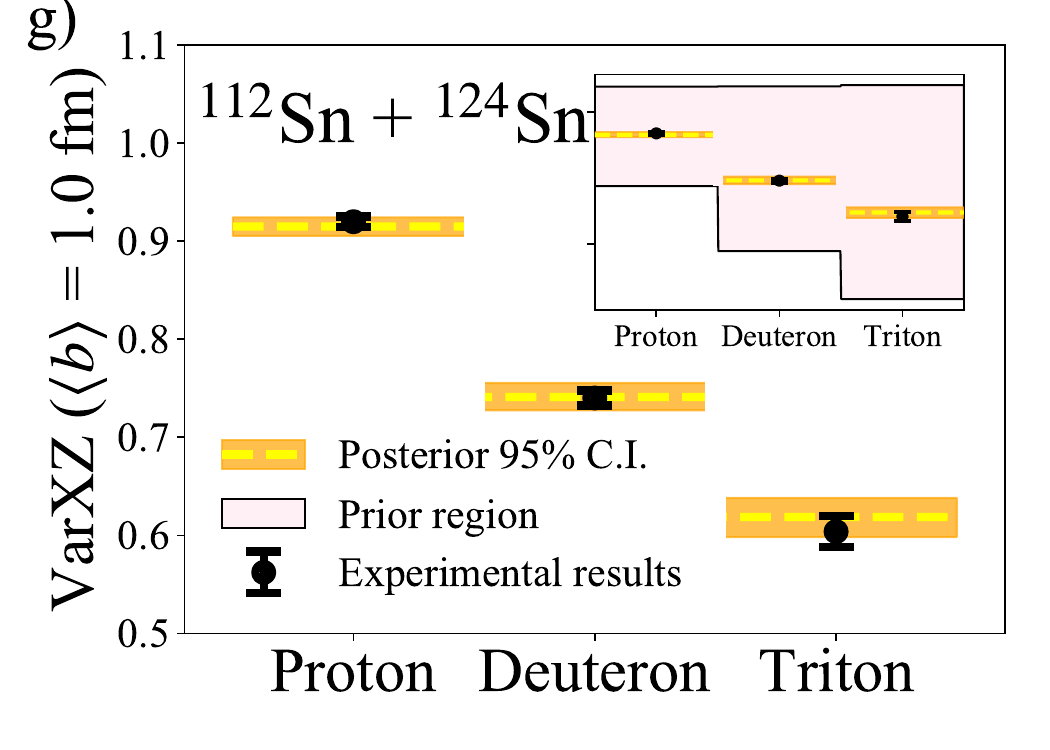}
    \end{subfigure}
    \caption{This figure compares directed (a, b, c, d), elliptic flow (e, f) and VarXZ (g) between the best fitted ImQMD-Sky predictions and experimental results. The black points show results from S$\pi$RIT experiment. The orange region shows the $2\sigma$ confidence region of ImQMD-Sky prediction from posterior distribution. The small sub-figure in each figure also shows a wide pink region that corresponds to the maximum ranges of prediction values from ImQMD-Sky with the parameter range in Table~\ref{tab:parRange}. (c) and (e) show results from the reaction system of (a) while (d) and (f) show results from the reaction systems of (b).}
    \label{fig:ExpFlowCompare}
\end{figure}

Figure~\ref{fig:ExpFlowCompare} shows the fitted flow and stopping observables. The first three rows show the results of directed and elliptic flow, with plots on the left column corresponding to the results of $^{108}$Sn$ + ^{112}$Sn reaction and the right $^{132}$Sn$ + ^{124}$Sn. From top to bottom, the three rows show $v_1$ as a function of $y_0$, $v_1$ as a function of $p_T$ (MeV) and $v_2$ as a function of $y_0$, all at $\langle b\rangle = \SI{5}{fm}$. The fourth row shows VarXZ for $^{112}$Sn$ + ^{124}$Sn at $\langle b\rangle = \SI{1}{fm}$.

\begin{figure}[h]
    \centering
    \includegraphics[width=\linewidth]{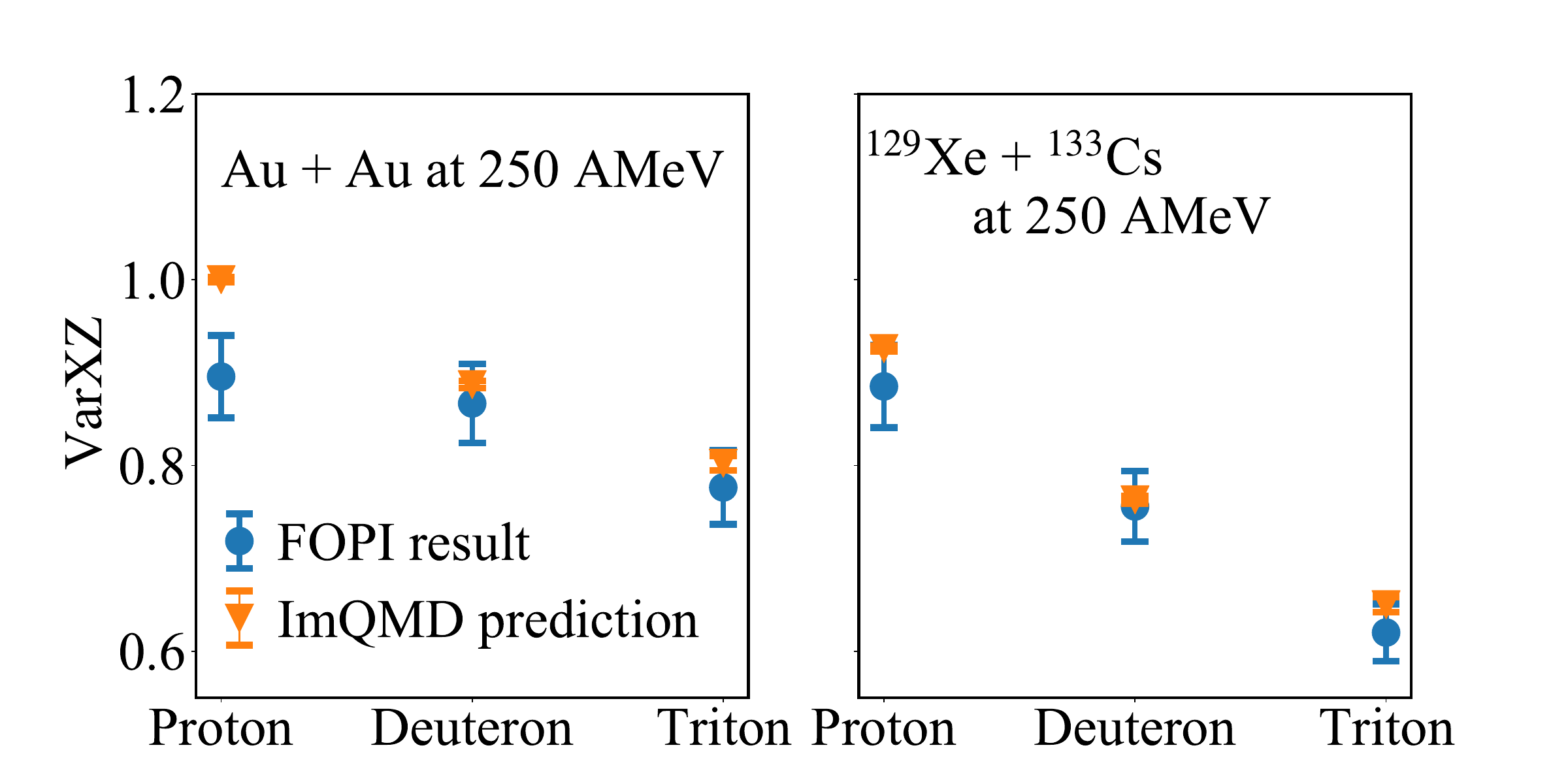}
    \caption{VarXZ of proton, deuteron and triton for $^{197}$Au + $^{197}$Au and $^{129}$Xe + $^{133}$Cs reactions at \SI{250}{AMeV} at $b = \SI{1}{fm}$. The orange inverted triangles  show ImQMD-Sky predictions using the best fitted parameter values obtained from the S$\pi$RIT experiment. The blue circles show experimental results from the FOPI data set.}
    \label{fig:FOPIVarXZ}
\end{figure}

To test the prediction power of our results, our most probable values of $S_0 = \SI{35}{MeV}$, $L = \SI{83}{MeV}$, $m_s^*/m_N = 0.84$, $m_v^*/m_N = 0.88$ and $\eta = - 0.01$ are used to predict VarXZ of protons, deuterons and tritons for $^{197}$Au + $^{197}$Au and $^{129}$Xe + $^{133}$Cs at a fixed-target beam energy of \SI{250}{AMeV} with the ImQMD-Sky model. VarXZs from these two systems are chosen because their values were measured experimentally in Ref.~\cite{REISDORF2010}. Experimental values of VarXZs at other beam energies are also available, but the beam energy at \SI{250}{AMeV} is the closest to the S$\pi$RIT beam energy of \SI{270}{AMeV}. Our choice minimizes effects due to changing beam energy and isolates the dependence on system size. Predictions from ImQMD-Sky  are plotted on top of experimental values in Fig.~\ref{fig:FOPIVarXZ} which shows reasonable agreement with an exception of proton from Au + Au at \SI{250}{AMeV}, where less stopping is observed than the model prediction. It is a good indication that our results are applicable to collisions of various system sizes near \SI{270}{AMeV}. 

The posterior on effective masses can be converted to a probability distribution on effective mass splitting $\Delta m_{np}^*/\delta$ using Eq.~\eqref{eq:mvConvert} and we find that $\Delta m_{np}^*/\delta = -0.07^{+0.07}_{-0.06}$. The value of effective mass splitting differs among various analyses using different data, but most of them favor a positive value~\cite{Li2022_2}. Of all the analyses in Ref.~\cite{Li2022_2}, the only result that favors a negative value of $\Delta m_{np}^*/\delta$ comes from the study of the n/p ratio from heavy ion reactions which yields  $\delta m_{np}^*/\delta = -0.05\pm0.09$~\cite{MORFOUACE2019}. That analysis also uses the ImQMD-Sky model for inference. It may indicates that the momentum dependence of the isovector mean fields in the ImQMD-Sky transport theory prefers a lower value of effective mass splitting. 

\begin{figure}[h]
    \centering
    \includegraphics[width=0.9\linewidth, trim={0 0.5cm 0 0.5cm},clip]{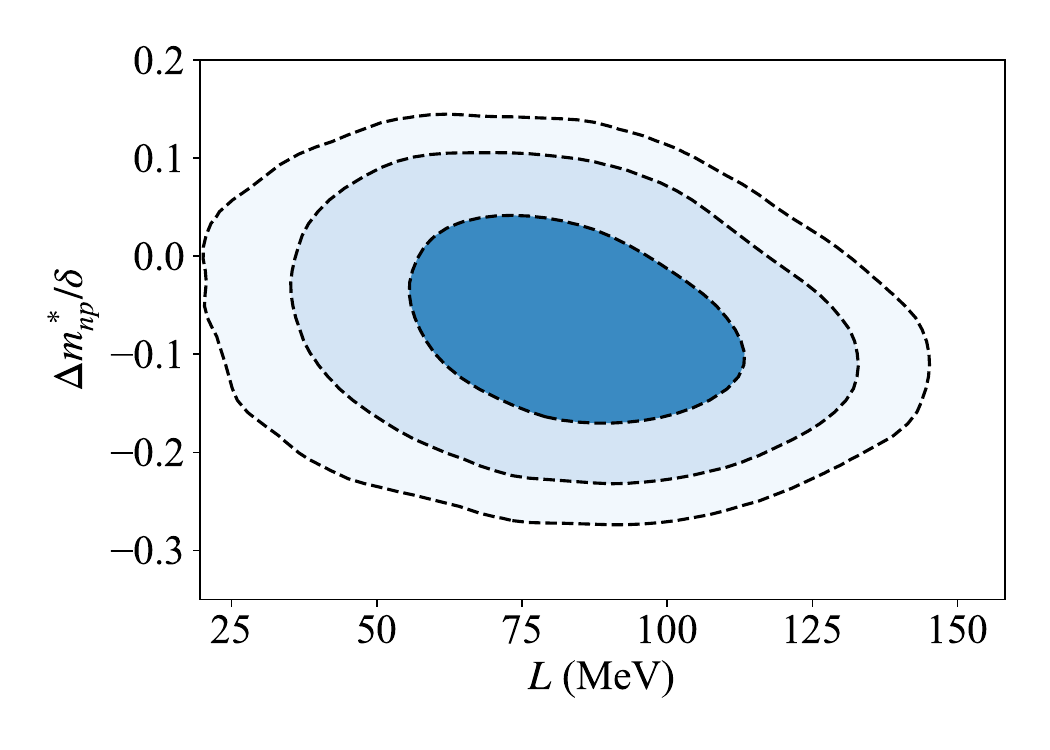}
    \caption{Correlation between $L$ and $\Delta m^*_{np}/\delta$. The three shades of blue, from the deepest to the lightest, correspond to 68\%, 95\% and 99\% confidence intervals. }
    \label{fig:dmnpL}
\end{figure}

Similar to constraints from pion observables~\cite{Estee2021}, we observed a correlation from this analysis between $\Delta m_{np}^*/\delta$ and $L$, as Fig.~\ref{fig:dmnpL} demonstrates.  The correlation trends in Ref.~\cite{Estee2021} are nearly orthogonal to the present work. While ImQMD-Sky is used for the current constraint, dcQMD was used in the pion analysis. Clearly, the model dependence of effective mass and symmetry energy effects must be studied carefully, ideally within an effort like the Transport Model Evaluation Project (TMEP)~\cite{Hermann2021}. Such endeavors will deepen our understanding of model dependence of EoS parameters and hopefully develop ways for constraints from different models to be compared and combined reliably. Even though a direct comparison is not feasible right now, this study opens an opportunity to improve constraints on $L$ by combining Fig.~\ref{fig:dmnpL} with correlated constraints between $L$ and $\Delta m_{np}^{*}/\delta$ from Ref.~\cite{Estee2021}.

\section{Summary and conclusion}
\label{sec:sum}

Directed flow ($v_1$) and elliptic flow ($v_2$) of $^{108}$Sn$ + ^{112}$Sn and $^{132}$Sn$ + ^{124}$Sn and stopping (VarXZ) of $^{112}$Sn$ + ^{124}$Sn (all at fixed-target beam energy of \SI{270}{AMeV}) are extracted from the data obtained in the S$\pi$RIT experiment. The measured values are compared to predictions from the Improved Quantum Molecular Dynamic-Skyrme (ImQMD-Sky) model through a Bayesian analysis which shows strong constraining power on the effective masses ($m_s^*/m_N$ and $m_v^*/m_N$) and the in-medium cross-section parameter ($\eta$). The most probable values are $m_s^*/m_N = 0.84\pm0.01$, $m_v^*/m_N=0.88\pm0.03$, and $\eta=-0.01\pm0.03$. The constraints on effective mass are converted to a probability distribution on effective mass splitting to give $\Delta m_{np}^*/\delta =-0.07^{+0.07}_{-0.06}$. This can be used to tighten constraints on symmetry energy terms such as $L$ as it was demonstrated to be correlated to $\Delta m_{np}^*/\delta$, but efforts in understanding model dependence of analysis on nucleon effective masses are warranted for the comparison to be conclusive.

\section{Acknowledgement}

The authors would like to thank members of the Transport Model Evaluation Project collaboration for many fruitful discussions. This work was supported by U.S. National Science Foundation Grant No. PHY-2209145, Department of Energy Grant No. DE-FG02-93ER40773, Department of Energy National Nuclear Security Administration Stewardship Science Graduate Fellowship under cooperative Agreement No. DE-NA0002135, the Robert A. Welch Foundation (A-1266 and A-1358), the Japanese MEXT, Japan KAKENHI (Grant-in-Aid for Scientific Research on Innovative Areas) grant No. 24105004, JSPS KAKENHI Grants Nos. JP17K05432, JP19K14709 and JP21K03528, the National Research Foundation of Korea under grant Nos. 2018R1A5A1025563 and 2013M7A1A1075764, the Polish National Science Center (NCN) under contract Nos. UMO-2013/09/B/ST2/04064, UMO-2013/-10/M/ST2/00624, the Ministry of Science and Technology of China under grant Nos. 2022YFE0103400 and Tsinghua University Initiative Scientific Research Program, Computing resources were provided by FRIB, the HOKUSAI-Great Wave system at RI-KEN, and the Institute for Cyber-Enabled Research at Michigan State University. 

\appendix
\section{S$\pi$RIT data}
\label{app:data}

All experimental data and uncertainties are tabulated in Table~\ref{tab:ExpValues}.
\renewcommand{\thetable}{A.\arabic{table}}
\setcounter{table}{0}
\begin{table}[!htbp]
\caption{Experimental values and the associated statistical and systematic uncertainties.}
\label{tab:ExpValues}
\begin{subtable}[t]{\linewidth}
\begin{center}
\caption{$v_1$ as a function of $y_0$ for $^{108}$Sn$ + ^{112}$Sn at $\langle b\rangle = \SI{5}{fm}$.}
\begin{tabular*}{\linewidth}{@{\extracolsep{\fill}}lllllll}
\toprule
\multicolumn{1}{c}{$y_0$} & 0.000 & 0.160 & 0.320 & 0.480 & 0.640 & 0.800 \\
\hline
\multicolumn{1}{c}{$v_1$} & -0.013 & 0.069 & 0.147 & 0.222 & 0.285 & 0.320 \\
Stat.                     & 0.002 & 0.002 & 0.002 & 0.002 & 0.002 & 0.002 \\
Sys.                      & 0.009 & 0.011 & 0.012 & 0.009 & 0.011 & 0.009 \\
\bottomrule
\end{tabular*}
\end{center}
\end{subtable}
\hspace{\fill}
\begin{subtable}[t]{\linewidth}
\begin{center}
\caption{$v_1$ as a function of $y_0$ for $^{132}$Sn$ + ^{124}$Sn at $\langle b\rangle = \SI{5}{fm}$.}
\begin{tabular*}{\linewidth}{@{\extracolsep{\fill}}lllllll}
\toprule
\multicolumn{1}{c}{$y_0$} & 0.000 & 0.160 & 0.320 & 0.480 & 0.640 & 0.800 \\
\hline
\multicolumn{1}{c}{$v_1$} & 0.020 & 0.102 & 0.184 & 0.259 & 0.316 & 0.337 \\
Stat                     & 0.002 & 0.002 & 0.002 & 0.002 & 0.002 & 0.002 \\
Sys                      & 0.007 & 0.008 & 0.009 & 0.009 & 0.008 & 0.007 \\
\bottomrule
\end{tabular*}
\end{center}
\end{subtable}

\begin{subtable}[t]{\linewidth}
\begin{center}
\caption{$v_1$ as a function of $p_T$ (MeV/c) for $^{108}$Sn$ + ^{112}$Sn at $\langle b\rangle = \SI{5}{fm}$.}
\begin{tabular*}{\linewidth}{@{\extracolsep{\fill}}llllll}
\toprule
\multicolumn{1}{c}{$p_T$} & 50 & 150 & 250 & 350 & 450 \\
\hline
\multicolumn{1}{c}{$v_1$}         & 0.136  & 0.283  & 0.293  & 0.257  & 0.233   \\
Stat.                             & 0.002  & 0.002   & 0.002 & 0.003   & 0.004   \\
Sys.                              & 0.005 & 0.010 & 0.016 & 0.011 & 0.013 \\
\bottomrule
\end{tabular*}
\end{center}
\end{subtable}
\hspace{\fill}
\begin{subtable}[t]{\linewidth}
\begin{center}
\caption{$v_1$ as a function of $p_T$ (MeV/c) for $^{132}$Sn$ + ^{124}$Sn at $\langle b\rangle = \SI{5}{fm}$.}
\begin{tabular*}{\linewidth}{@{\extracolsep{\fill}}llllll}
\toprule
\multicolumn{1}{c}{$p_T$} & 50 & 150 & 250 & 350 & 450 \\
\hline
\multicolumn{1}{c}{$v_1$} & 0.152 & 0.322 & 0.327 & 0.284 & 0.258 \\
Stat & 0.002 & 0.002 & 0.002 & 0.003 & 0.004 \\
Sys & 0.004 & 0.009 & 0.013 & 0.013 & 0.013 \\
\bottomrule
\end{tabular*}
\end{center}
\end{subtable}

\begin{subtable}[t]{\linewidth}
\begin{center}
\caption{$v_2$ as a function of $y_0$ for $^{108}$Sn$ + ^{112}$Sn at $\langle b\rangle = \SI{5}{fm}$.}
\begin{tabular*}{\linewidth}{@{\extracolsep{\fill}}lllllll}
\toprule
\multicolumn{1}{c}{$y_0$} & 0.000  & 0.160  & 0.320  & 0.480  & 0.640  & 0.800  \\
\hline
\multicolumn{1}{c}{$v_2$} & -0.049  & -0.051  & -0.044  & -0.031  & -0.019  & -0.006 \\
Stat.                     & 0.003  & 0.003  & 0.003  & 0.003  & 0.003  & 0.003  \\
Sys.                      & 0.006 &	0.006 &	0.010 &	0.010 &	0.013 &	0.014 \\
\bottomrule
\end{tabular*}
\end{center}
\end{subtable}
\hspace{\fill}
\begin{subtable}[t]{\linewidth}
\begin{center}
\caption{$v_2$ as a function of $y_0$ for $^{132}$Sn$ + ^{124}$Sn at $\langle b\rangle = \SI{5}{fm}$.}
\begin{tabular*}{\linewidth}{@{\extracolsep{\fill}}lllllll}
\toprule
\multicolumn{1}{c}{$y_0$} & 0.000  & 0.160  & 0.320  & 0.480  & 0.640  & 0.800  \\
\hline
\multicolumn{1}{c}{$v_2$} & -0.058  & -0.049  & -0.038  & -0.023  & -0.006  & -0.003 \\
Stat.                     & 0.003  & 0.003  & 0.003  & 0.003  & 0.003  & 0.003  \\
Sys.                      & 0.008  &  0.007 & 0.007 & 0.011 & 0.014 & 0.012 \\
\bottomrule
\end{tabular*}
\end{center}
\end{subtable}

\begin{subtable}[t]{\linewidth}
\begin{center}
\caption{VarXZ of proton, deuteron and triton for $^{112}$Sn$ + ^{124}$Sn at $\langle b\rangle = \SI{1}{fm}$.}
\begin{tabular*}{\linewidth}{@{\extracolsep{\fill}}llll}
\toprule
\multicolumn{1}{c}{$y_0$} & proton & deuteron & triton  \\
\hline
\multicolumn{1}{c}{VarXZ} &  0.915 & 0.767 & 0.590 \\
Stat.                     &  0.004 & 0.005 & 0.003 \\
Sys.                      &  0.011 & 0.012 & 0.010\\
\bottomrule
\end{tabular*}
\end{center}
\end{subtable}
\end{table}

\begin{thebibliography}{10}
\expandafter\ifx\csname url\endcsname\relax
  \def\url#1{\texttt{#1}}\fi
\expandafter\ifx\csname urlprefix\endcsname\relax\def\urlprefix{URL }\fi
\expandafter\ifx\csname href\endcsname\relax
  \def\href#1#2{#2} \def\path#1{#1}\fi

\bibitem{Xu2019}
J.~Xu,
  \href{https://www.sciencedirect.com/science/article/pii/S0146641019300213}{Transport
  approaches for the description of intermediate-energy heavy-ion collisions},
  Progress in Particle and Nuclear Physics 106 (2019) 312--359.
\newblock \href {https://doi.org/https://doi.org/10.1016/j.ppnp.2019.02.009}
  {\path{doi:https://doi.org/10.1016/j.ppnp.2019.02.009}}.
\newline\urlprefix\url{https://www.sciencedirect.com/science/article/pii/S0146641019300213}

\bibitem{sorensen2023}
A.~Sorensen, K.~Agarwal, K.~W. Brown, Z.~Chajecki, P.~Danielewicz,
  C.~Drischler, S.~Gandolfi, J.~W. Holt, M.~Kaminski, C.-M. Ko, R.~Kumar, B.-A.
  Li, W.~G. Lynch, A.~B. McIntosh, W.~G. Newton, S.~Pratt, O.~Savchuk,
  M.~Stefaniak, I.~Tews, M.~B. Tsang, R.~Vogt, H.~Wolter, H.~Zbroszczyk,
  N.~Abbasi, J.~Aichelin, A.~Andronic, S.~A. Bass, F.~Becattini, D.~Blaschke,
  M.~Bleicher, C.~Blume, E.~Bratkovskaya, B.~A. Brown, D.~A. Brown,
  A.~Camaiani, G.~Casini, K.~Chatziioannou, A.~Chbihi, M.~Colonna, M.~D. Cozma,
  V.~Dexheimer, X.~Dong, T.~Dore, L.~Du, J.~A. Dueñas, H.~Elfner,
  W.~Florkowski, Y.~Fujimoto, R.~J. Furnstahl, A.~Gade, T.~Galatyuk, C.~Gale,
  F.~Geurts, S.~Grozdanov, K.~Hagel, S.~P. Harris, W.~Haxton, U.~Heinz, M.~P.
  Heller, O.~Hen, H.~Hergert, N.~Herrmann, H.~Z. Huang, X.-G. Huang, N.~Ikeno,
  G.~Inghirami, J.~Jankowski, J.~Jia, J.~C. Jiménez, J.~Kapusta, B.~Kardan,
  I.~Karpenko, D.~Keane, D.~Kharzeev, A.~Kugler, A.~L. Fèvre, D.~Lee, H.~Liu,
  M.~A. Lisa, W.~J. Llope, I.~Lombardo, M.~Lorenz, T.~Marchi, L.~McLerran,
  U.~Mosel, A.~Motornenko, B.~Müller, P.~Napolitani, J.~B. Natowitz,
  W.~Nazarewicz, J.~Noronha, J.~Noronha-Hostler, G.~Odyniec,
  P.~Papakonstantinou, Z.~Paulínyová, J.~Piekarewicz, R.~D. Pisarski,
  C.~Plumberg, M.~Prakash, J.~Randrup, C.~Ratti, P.~Rau, S.~Reddy, H.-R.
  Schmidt, P.~Russotto, R.~Ryblewski, A.~Schäfer, B.~Schenke, S.~Sen,
  P.~Senger, R.~Seto, C.~Shen, B.~Sherrill, M.~Singh, V.~Skokov, M.~Spaliński,
  J.~Steinheimer, M.~Stephanov, J.~Stroth, C.~Sturm, K.-J. Sun, A.~Tang,
  G.~Torrieri, W.~Trautmann, G.~Verde, V.~Vovchenko, R.~Wada, F.~Wang, G.~Wang,
  K.~Werner, N.~Xu, Z.~Xu, H.-U. Yee, S.~Yennello, Y.~Yin, Dense nuclear matter
  equation of state from heavy-ion collisions (2023).
\newblock \href {http://arxiv.org/abs/2301.13253} {\path{arXiv:2301.13253}}.

\bibitem{LYNCH2022137098}
W.~Lynch, M.~Tsang,
  \href{https://www.sciencedirect.com/science/article/pii/S0370269322002325}{Decoding
  the density dependence of the nuclear symmetry energy}, Physics Letters B 830
  (2022) 137098.
\newblock \href
  {https://doi.org/https://doi.org/10.1016/j.physletb.2022.137098}
  {\path{doi:https://doi.org/10.1016/j.physletb.2022.137098}}.
\newline\urlprefix\url{https://www.sciencedirect.com/science/article/pii/S0370269322002325}

\bibitem{Danielewicz2002}
P.~Danielewicz, R.~Lacey, W.~G. Lynch, {D}etermination of the {E}quation of
  {S}tate of {D}ense {M}atter, Science 298~(5598) (2002) 1592--1596.
\newblock \href {https://doi.org/10.1126/science.1078070}
  {\path{doi:10.1126/science.1078070}}.

\bibitem{LEFEVRE2016}
A.~L. F{\`e}vre, Y.~Leifels, W.~Reisdorf, J.~Aichelin, C.~Hartnack,
  \href{http://www.sciencedirect.com/science/article/pii/S0375947415002225}{Constraining
  the nuclear matter equation of state around twice saturation density},
  Nuclear Physics A 945 (2016) 112 -- 133.
\newblock \href
  {https://doi.org/https://doi.org/10.1016/j.nuclphysa.2015.09.015}
  {\path{doi:https://doi.org/10.1016/j.nuclphysa.2015.09.015}}.
\newline\urlprefix\url{http://www.sciencedirect.com/science/article/pii/S0375947415002225}

\bibitem{Dutra2012}
M.~Dutra, O.~Louren\ifmmode~\mbox{\c{c}}\else \c{c}\fi{}o, J.~S. S\'a~Martins,
  A.~Delfino, J.~R. Stone, P.~D. Stevenson,
  \href{https://link.aps.org/doi/10.1103/PhysRevC.85.035201}{{S}kyrme
  interaction and nuclear matter constraints}, {P}hys. {R}ev. {C} 85 (2012)
  035201.
\newblock \href {https://doi.org/10.1103/PhysRevC.85.035201}
  {\path{doi:10.1103/PhysRevC.85.035201}}.
\newline\urlprefix\url{https://link.aps.org/doi/10.1103/PhysRevC.85.035201}

\bibitem{zhang2015electric}
Z.~Zhang, L.-W. Chen, {Electric dipole polarizability in $^{208}$Pb as a probe
  of the symmetry energy and neutron matter around $\rho_0/3$}, Physical Review
  C 92~(3) (2015) 031301.

\bibitem{PREX2021}
D.~Adhikari, H.~Albataineh, D.~Androic, K.~Aniol, D.~S. Armstrong, T.~Averett,
  C.~Ayerbe~Gayoso, S.~Barcus, V.~Bellini, R.~S. Beminiwattha, J.~F. Benesch,
  H.~Bhatt, D.~Bhatta~Pathak, D.~Bhetuwal, B.~Blaikie, Q.~Campagna,
  A.~Camsonne, G.~D. Cates, Y.~Chen, C.~Clarke, J.~C. Cornejo, S.~Covrig~Dusa,
  P.~Datta, A.~Deshpande, D.~Dutta, C.~Feldman, E.~Fuchey, C.~Gal, D.~Gaskell,
  T.~Gautam, M.~Gericke, C.~Ghosh, I.~Halilovic, J.-O. Hansen, F.~Hauenstein,
  W.~Henry, C.~J. Horowitz, C.~Jantzi, S.~Jian, S.~Johnston, D.~C. Jones,
  B.~Karki, S.~Katugampola, C.~Keppel, P.~M. King, D.~E. King, M.~Knauss, K.~S.
  Kumar, T.~Kutz, N.~Lashley-Colthirst, G.~Leverick, H.~Liu, N.~Liyange,
  S.~Malace, R.~Mammei, J.~Mammei, M.~McCaughan, D.~McNulty, D.~Meekins,
  C.~Metts, R.~Michaels, M.~M. Mondal, J.~Napolitano, A.~Narayan, D.~Nikolaev,
  M.~N.~H. Rashad, V.~Owen, C.~Palatchi, J.~Pan, B.~Pandey, S.~Park, K.~D.
  Paschke, M.~Petrusky, M.~L. Pitt, S.~Premathilake, A.~J.~R. Puckett,
  B.~Quinn, R.~Radloff, S.~Rahman, A.~Rathnayake, B.~T. Reed, P.~E. Reimer,
  R.~Richards, S.~Riordan, Y.~Roblin, S.~Seeds, A.~Shahinyan, P.~Souder,
  L.~Tang, M.~Thiel, Y.~Tian, G.~M. Urciuoli, E.~W. Wertz, B.~Wojtsekhowski,
  B.~Yale, T.~Ye, A.~Yoon, A.~Zec, W.~Zhang, J.~Zhang, X.~Zheng,
  \href{https://link.aps.org/doi/10.1103/PhysRevLett.126.172502}{{A}ccurate
  {D}etermination of the {N}eutron {S}kin {T}hickness of $^{208}\mathrm{{P}b}$
  through {P}arity-{V}iolation in {E}lectron {S}cattering}, {P}hys. {R}ev.
  {L}ett. 126 (2021) 172502.
\newblock \href {https://doi.org/10.1103/PhysRevLett.126.172502}
  {\path{doi:10.1103/PhysRevLett.126.172502}}.
\newline\urlprefix\url{https://link.aps.org/doi/10.1103/PhysRevLett.126.172502}

\bibitem{reed2021implications}
B.~T. Reed, F.~J. Fattoyev, C.~J. Horowitz, J.~Piekarewicz, {Implications of
  PREX-2 on the Equation of State of Neutron-Rich Matter}, Physical Review
  Letters 126~(17) (2021) 172503.

\bibitem{brown2013}
B.~A. Brown,
  \href{https://link.aps.org/doi/10.1103/PhysRevLett.111.232502}{{C}onstraints
  on the {S}kyrme {E}quations of {S}tate from {P}roperties of {D}oubly {M}agic
  {N}uclei}, {P}hys. {R}ev. {L}ett. 111 (2013) 232502.
\newblock \href {https://doi.org/10.1103/PhysRevLett.111.232502}
  {\path{doi:10.1103/PhysRevLett.111.232502}}.
\newline\urlprefix\url{https://link.aps.org/doi/10.1103/PhysRevLett.111.232502}

\bibitem{kortelainen2012nuclear}
M.~Kortelainen, J.~McDonnell, W.~Nazarewicz, P.-G. Reinhard, J.~Sarich,
  N.~Schunck, M.~V. Stoitsov, S.~M. Wild, {Nuclear energy density optimization:
  Large deformations}, Physical Review C 85~(2) (2012) 024304.

\bibitem{Danielewicz2016}
P.~Danielewicz, P.~Singh, J.~Lee, {Symmetry Energy III: Isovector Skins}, Nucl.
  Phys. A958 (2017) 147--186.
\newblock \href {http://arxiv.org/abs/1611.01871} {\path{arXiv:1611.01871}},
  \href {https://doi.org/10.1016/j.nuclphysa.2016.11.008}
  {\path{doi:10.1016/j.nuclphysa.2016.11.008}}.

\bibitem{tsang2009}
M.~B. Tsang, Y.~Zhang, P.~Danielewicz, M.~Famiano, Z.~Li, W.~G. Lynch, A.~W.
  Steiner,
  \href{https://link.aps.org/doi/10.1103/PhysRevLett.102.122701}{{C}onstraints
  on the {D}ensity {D}ependence of the {S}ymmetry {E}nergy}, {P}hys. {R}ev.
  {L}ett. 102 (2009) 122701.
\newblock \href {https://doi.org/10.1103/PhysRevLett.102.122701}
  {\path{doi:10.1103/PhysRevLett.102.122701}}.
\newline\urlprefix\url{https://link.aps.org/doi/10.1103/PhysRevLett.102.122701}

\bibitem{MORFOUACE2019}
P.~Morfouace, C.~Tsang, Y.~Zhang, W.~Lynch, M.~Tsang, D.~Coupland, M.~Youngs,
  Z.~Chajecki, M.~Famiano, T.~Ghosh, G.~Jhang, J.~Lee, H.~Liu, A.~Sanetullaev,
  R.~Showalter, J.~Winkelbauer,
  \href{https://www.sciencedirect.com/science/article/pii/S0370269319307671}{{C}onstraining
  the symmetry energy with heavy-ion collisions and {B}ayesian analyses},
  {P}hysics {L}etters {B} 799 (2019) 135045.
\newblock \href
  {https://doi.org/https://doi.org/10.1016/j.physletb.2019.135045}
  {\path{doi:https://doi.org/10.1016/j.physletb.2019.135045}}.
\newline\urlprefix\url{https://www.sciencedirect.com/science/article/pii/S0370269319307671}

\bibitem{Estee2021}
J.~Estee, W.~G. Lynch, C.~Y. Tsang, J.~Barney, G.~Jhang, M.~B. Tsang, R.~Wang,
  M.~Kaneko, J.~W. Lee, T.~Isobe, M.~Kurata-Nishimura, T.~Murakami, D.~S. Ahn,
  L.~Atar, T.~Aumann, H.~Baba, K.~Boretzky, J.~Brzychczyk, G.~Cerizza,
  N.~Chiga, N.~Fukuda, I.~Gasparic, B.~Hong, A.~Horvat, K.~Ieki, N.~Inabe,
  Y.~J. Kim, T.~Kobayashi, Y.~Kondo, P.~Lasko, H.~S. Lee, Y.~Leifels,
  J.~\L{}ukasik, J.~Manfredi, A.~B. McIntosh, P.~Morfouace, T.~Nakamura,
  N.~Nakatsuka, S.~Nishimura, H.~Otsu, P.~Paw\l{}owski, K.~Pelczar, D.~Rossi,
  H.~Sakurai, C.~Santamaria, H.~Sato, H.~Scheit, R.~Shane, Y.~Shimizu,
  H.~Simon, A.~Snoch, A.~Sochocka, T.~Sumikama, H.~Suzuki, D.~Suzuki,
  H.~Takeda, S.~Tangwancharoen, H.~Toernqvist, Y.~Togano, Z.~G. Xiao, S.~J.
  Yennello, Y.~Zhang, M.~D. Cozma,
  \href{https://link.aps.org/doi/10.1103/PhysRevLett.126.162701}{{P}robing the
  {S}ymmetry {E}nergy with the {S}pectral {P}ion {R}atio}, {P}hys. {R}ev.
  {L}ett. 126 (2021) 162701.
\newblock \href {https://doi.org/10.1103/PhysRevLett.126.162701}
  {\path{doi:10.1103/PhysRevLett.126.162701}}.
\newline\urlprefix\url{https://link.aps.org/doi/10.1103/PhysRevLett.126.162701}

\bibitem{cozma2018feasibility}
M.~Cozma, Feasibility of constraining the curvature parameter of the symmetry
  energy using elliptic flow data, The European Physical Journal A 54~(3)
  (2018) 1--23.

\bibitem{RUSSOTTO2011471}
P.~Russotto, P.~Wu, M.~Zoric, M.~Chartier, Y.~Leifels, R.~Lemmon, Q.~Li,
  J.~Łukasik, A.~Pagano, P.~Pawłowski, W.~Trautmann,
  \href{https://www.sciencedirect.com/science/article/pii/S037026931100178X}{Symmetry
  energy from elliptic flow in 197au+197au}, Physics Letters B 697~(5) (2011)
  471--476.
\newblock \href
  {https://doi.org/https://doi.org/10.1016/j.physletb.2011.02.033}
  {\path{doi:https://doi.org/10.1016/j.physletb.2011.02.033}}.
\newline\urlprefix\url{https://www.sciencedirect.com/science/article/pii/S037026931100178X}

\bibitem{Russotto2016}
P.~Russotto, et~al.,
  \href{https://link.aps.org/doi/10.1103/PhysRevC.94.034608}{{R}esults of the
  {A}{S}{Y}-{E}{O}{S} experiment at {G}{S}{I}: {T}he symmetry energy at
  suprasaturation density}, {P}hys. {R}ev. {C} 94 (2016) 034608.
\newblock \href {https://doi.org/10.1103/PhysRevC.94.034608}
  {\path{doi:10.1103/PhysRevC.94.034608}}.
\newline\urlprefix\url{https://link.aps.org/doi/10.1103/PhysRevC.94.034608}

\bibitem{Margueron2018}
J.~Margueron, R.~Hoffmann~Casali, F.~Gulminelli,
  \href{https://link.aps.org/doi/10.1103/PhysRevC.97.025805}{{E}quation of
  state for dense nucleonic matter from metamodeling. {I}. {F}oundational
  aspects}, {P}hys. {R}ev. {C} 97 (2018) 025805.
\newblock \href {https://doi.org/10.1103/PhysRevC.97.025805}
  {\path{doi:10.1103/PhysRevC.97.025805}}.
\newline\urlprefix\url{https://link.aps.org/doi/10.1103/PhysRevC.97.025805}

\bibitem{Zhang20202}
Y.~Zhang, M.~Liu, C.-J. Xia, Z.~Li, S.~K. Biswal,
  \href{https://link.aps.org/doi/10.1103/PhysRevC.101.034303}{Constraints on
  the symmetry energy and its associated parameters from nuclei to neutron
  stars}, Phys. Rev. C 101 (2020) 034303.
\newblock \href {https://doi.org/10.1103/PhysRevC.101.034303}
  {\path{doi:10.1103/PhysRevC.101.034303}}.
\newline\urlprefix\url{https://link.aps.org/doi/10.1103/PhysRevC.101.034303}

\bibitem{Li2004}
B.-A. Li,
  \href{https://link.aps.org/doi/10.1103/PhysRevC.69.064602}{{C}onstraining the
  neutron-proton effective mass splitting in neutron-rich matter}, {P}hys.
  {R}ev. {C} 69 (2004) 064602.
\newblock \href {https://doi.org/10.1103/PhysRevC.69.064602}
  {\path{doi:10.1103/PhysRevC.69.064602}}.
\newline\urlprefix\url{https://link.aps.org/doi/10.1103/PhysRevC.69.064602}

\bibitem{Brueckner1955}
K.~Brueckner,
  \href{https://www.scopus.com/inward/record.uri?eid=2-s2.0-36149019045&doi=10.1103%2fPhysRev.97.1353&partnerID=40&md5=e718dea1d3cc1243be7b9ecff14acd94}{{T}wo-body
  forces and nuclear saturation. {I}{I}{I}. {D}etails of the structure of the
  nucleus}, {P}hysical {R}eview 97~(5) (1955) 1353--1366, cited By 388.
\newblock \href {https://doi.org/10.1103/PhysRev.97.1353}
  {\path{doi:10.1103/PhysRev.97.1353}}.
\newline\urlprefix\url{https://www.scopus.com/inward/record.uri?eid=2-s2.0-36149019045&doi=10.1103%2fPhysRev.97.1353&partnerID=40&md5=e718dea1d3cc1243be7b9ecff14acd94}

\bibitem{MAHAUX1985}
C.~Mahaux, P.~Bortignon, R.~Broglia, C.~Dasso,
  \href{https://www.sciencedirect.com/science/article/pii/0370157385901000}{{D}ynamics
  of the shell model}, {P}hysics {R}eports 120~(1) (1985) 1--274.
\newblock \href {https://doi.org/https://doi.org/10.1016/0370-1573(85)90100-0}
  {\path{doi:https://doi.org/10.1016/0370-1573(85)90100-0}}.
\newline\urlprefix\url{https://www.sciencedirect.com/science/article/pii/0370157385901000}

\bibitem{Ring1980}
P.~Ring, P.~Schuck, The nuclear many-body problem, Springer-Verlag, New York,
  1980.

\bibitem{li2018}
B.-A. Li, B.-J. Cai, L.-W. Chen, J.~Xu, Nucleon effective masses in
  neutron-rich matter, Progress in Particle and Nuclear Physics 99 (2018)
  29--119.

\bibitem{Barney2019}
J.~E. Barney, {{C}harged {P}ion {E}mission from $^{112}${S}n + $^{124}${S}n and
  $^{124}${S}n + $^{112}${S}n {R}eactions with the {S}$\pi${R}{I}{T} {T}ime
  {P}rojection {C}hamber}, Ph.D. thesis.

\bibitem{Shane2015}
R.~Shane, A.~B. McIntosh, T.~Isobe, W.~G. Lynch, H.~Baba, J.~Barney,
  Z.~Chajecki, M.~Chartier, J.~Estee, M.~Famiano, B.~Hong, K.~Ieki, G.~Jhang,
  R.~Lemmon, F.~Lu, T.~Murakami, N.~Nakatsuka, M.~Nishimura, R.~Olsen,
  W.~Powell, H.~Sakurai, A.~Taketani, S.~Tangwancharoen, M.~B. Tsang,
  T.~Usukura, R.~Wang, S.~J. Yennello, J.~Yurkon,
  \href{http://www.sciencedirect.com/science/article/pii/S0168900215000534}{S$\pi$rit:
  A time-projection chamber for symmetry-energy studies}, Nuclear Instruments
  and Methods in Physics Research Section A: Accelerators, Spectrometers,
  Detectors and Associated Equipment 784 (2015) 513 -- 517, {S}ymposium on
  Radiation Measurements and Applications 2014 (SORMA XV).
\newblock \href {https://doi.org/https://doi.org/10.1016/j.nima.2015.01.026}
  {\path{doi:https://doi.org/10.1016/j.nima.2015.01.026}}.
\newline\urlprefix\url{http://www.sciencedirect.com/science/article/pii/S0168900215000534}

\bibitem{estee2019}
J.~Estee, W.~Lynch, J.~Barney, G.~Cerizza, G.~Jhang, J.~Lee, R.~Wang, T.~Isobe,
  M.~Kaneko, M.~Kurata-Nishimura, T.~Murakami, R.~Shane, S.~Tangwancharoen,
  C.~Tsang, M.~Tsang, B.~Hong, P.~Lasko, J.~Łukasik, A.~McIntosh,
  P.~Pawłowski, K.~Pelczar, H.~Sakurai, C.~Santamaria, D.~Suzuki, S.~Yennello,
  Y.~Zhang,
  \href{https://www.sciencedirect.com/science/article/pii/S0168900219310472}{{E}xtending
  the dynamic range of electronics in a {T}ime {P}rojection {C}hamber},
  {N}uclear {I}nstruments and {M}ethods in {P}hysics {R}esearch {S}ection {A}:
  {A}ccelerators, {S}pectrometers, {D}etectors and {A}ssociated {E}quipment 944
  (2019) 162509.
\newblock \href {https://doi.org/https://doi.org/10.1016/j.nima.2019.162509}
  {\path{doi:https://doi.org/10.1016/j.nima.2019.162509}}.
\newline\urlprefix\url{https://www.sciencedirect.com/science/article/pii/S0168900219310472}

\bibitem{LEE2020}
J.~Lee, G.~Jhang, G.~Cerizza, J.~Barney, J.~Estee, T.~Isobe, M.~Kaneko,
  M.~Kurata-Nishimura, W.~Lynch, T.~Murakami, C.~Tsang, M.~Tsang, R.~Wang,
  B.~Hong, A.~McIntosh, H.~Sakurai, C.~Santamaria, R.~Shane, S.~Tangwancharoen,
  S.~Yennello, Y.~Zhang,
  \href{https://www.sciencedirect.com/science/article/pii/S0168900220303545}{{C}harged
  particle track reconstruction with {S}$\pi${R}{I}{T} {T}ime {P}rojection
  {C}hamber}, {N}uclear {I}nstruments and {M}ethods in {P}hysics {R}esearch
  {S}ection {A}: {A}ccelerators, {S}pectrometers, {D}etectors and {A}ssociated
  {E}quipment 965 (2020) 163840.
\newblock \href {https://doi.org/https://doi.org/10.1016/j.nima.2020.163840}
  {\path{doi:https://doi.org/10.1016/j.nima.2020.163840}}.
\newline\urlprefix\url{https://www.sciencedirect.com/science/article/pii/S0168900220303545}

\bibitem{JHANG2021}
G.~Jhang, J.~Estee, J.~Barney, G.~Cerizza, M.~Kaneko, J.~Lee, W.~Lynch,
  T.~Isobe, M.~Kurata-Nishimura, T.~Murakami, C.~Tsang, M.~Tsang, R.~Wang,
  D.~Ahn, L.~Atar, T.~Aumann, H.~Baba, K.~Boretzky, J.~Brzychczyk, N.~Chiga,
  N.~Fukuda, I.~Gasparic, B.~Hong, A.~Horvat, K.~Ieki, N.~Inabe, Y.~Kim,
  T.~Kobayashi, Y.~Kondo, P.~Lasko, H.~Lee, Y.~Leifels, J.~Łukasik,
  J.~Manfredi, A.~McIntosh, P.~Morfouace, T.~Nakamura, N.~Nakatsuka,
  S.~Nishimura, R.~Olsen, H.~Otsu, P.~Pawłowski, K.~Pelczar, D.~Rossi,
  H.~Sakurai, C.~Santamaria, H.~Sato, H.~Scheit, R.~Shane, Y.~Shimizu,
  H.~Simon, A.~Snoch, A.~Sochocka, Z.~Sosin, T.~Sumikama, H.~Suzuki, D.~Suzuki,
  H.~Takeda, S.~Tangwancharoen, H.~Toernqvist, Y.~Togano, Z.~Xiao, S.~Yennello,
  J.~Yurkon, Y.~Zhang, M.~Colonna, D.~Cozma, P.~Danielewicz, H.~Elfner,
  N.~Ikeno, C.~M. Ko, J.~Mohs, D.~Oliinychenko, A.~Ono, J.~Su, Y.~J. Wang,
  H.~Wolter, J.~Xu, Y.-X. Zhang, Z.~Zhang,
  \href{https://www.sciencedirect.com/science/article/pii/S0370269320308194}{{S}ymmetry
  energy investigation with pion production from {S}n+{S}n systems}, {P}hysics
  {L}etters {B} 813 (2021) 136016.
\newblock \href
  {https://doi.org/https://doi.org/10.1016/j.physletb.2020.136016}
  {\path{doi:https://doi.org/10.1016/j.physletb.2020.136016}}.
\newline\urlprefix\url{https://www.sciencedirect.com/science/article/pii/S0370269320308194}

\bibitem{Masanori2022}
M.~Kaneko, {Hydrogen Isotope Productions in Sn + Sn Collisions with Radioactive
  Beams at 270 MeV/nucleon}, Ph.D. thesis, Kyoto University (2022).

\bibitem{Lee2022}
J.~W. Lee, M.~B. Tsang, C.~Y. Tsang, R.~Wang, J.~Barney, J.~Estee, T.~Isobe,
  M.~Kaneko, M.~Kurata-Nishimura, W.~G. Lynch, T.~Murakami, A.~Ono, S.~R.
  Souza, D.~S. Ahn, L.~Atar, T.~Aumann, H.~Baba, K.~Boretzky, J.~Brzychczyk,
  G.~Cerizza, N.~Chiga, N.~Fukuda, I.~Gasparic, B.~Hong, A.~Horvat, K.~Ieki,
  N.~Ikeno, N.~Inabe, G.~Jhang, Y.~J. Kim, T.~Kobayashi, Y.~Kondo, P.~Lasko,
  H.~S. Lee, Y.~Leifels, J.~{\L}ukasik, J.~Manfredi, A.~B. McIntosh,
  P.~Morfouace, T.~Nakamura, N.~Nakatsuka, S.~Nishimura, H.~Otsu,
  P.~Paw{\l}owski, K.~Pelczar, D.~Rossi, H.~Sakurai, C.~Santamaria, H.~Sato,
  H.~Scheit, R.~Shane, Y.~Shimizu, H.~Simon, A.~Snoch, A.~Sochocka,
  T.~Sumikama, H.~Suzuki, D.~Suzuki, H.~Takeda, S.~Tangwancharoen, Y.~Togano,
  Z.~G. Xiao, S.~J. Yennello, Y.~Zhang, t.~S. p.~R. collaboration),
  \href{https://doi.org/10.1140/epja/s10050-022-00851-2}{{Isoscaling in central
  Sn+Sn collisions at 270 MeV/u}}, The European Physical Journal A 58~(10)
  (2022) 201.
\newblock \href {https://doi.org/10.1140/epja/s10050-022-00851-2}
  {\path{doi:10.1140/epja/s10050-022-00851-2}}.
\newline\urlprefix\url{https://doi.org/10.1140/epja/s10050-022-00851-2}

\bibitem{RAMI1999}
F.~Rami, P.~Crochet, R.~Donà, B.~{de Schauenburg}, P.~Wagner, J.~Alard,
  A.~Andronic, Z.~Basrak, N.~Bastid, I.~Belyaev, A.~Bendarag, G.~Berek,
  D.~Best, R.~Čaplar, A.~Devismes, P.~Dupieux, M.~Dželalija, M.~Eskef,
  Z.~Fodor, A.~Gobbi, Y.~Grishkin, N.~Herrmann, K.~Hildenbrand, B.~Hong,
  J.~Kecskemeti, M.~Kirejczyk, M.~Korolija, R.~Kotte, A.~Lebedev, Y.~Leifels,
  H.~Merlitz, S.~Mohren, D.~Moisa, W.~Neubert, D.~Pelte, M.~Petrovici,
  C.~Pinkenburg, C.~Plettner, W.~Reisdorf, D.~Schüll, Z.~Seres, B.~Sikora,
  V.~Simion, K.~Siwek-Wilczyńska, G.~Stoicea, M.~Stockmeir, M.~Vasiliev,
  K.~Wisniewski, D.~Wohlfarth, I.~Yushmanov, A.~Zhilin,
  \href{https://www.sciencedirect.com/science/article/pii/S0375947498006411}{{F}low
  angle from intermediate mass fragment measurements}, {N}uclear {P}hysics {A}
  646~(3) (1999) 367--384.
\newblock \href {https://doi.org/https://doi.org/10.1016/S0375-9474(98)00641-1}
  {\path{doi:https://doi.org/10.1016/S0375-9474(98)00641-1}}.
\newline\urlprefix\url{https://www.sciencedirect.com/science/article/pii/S0375947498006411}

\bibitem{Andronic2001}
A.~Andronic, W.~Reisdorf, J.~P.~Alard, V.~Barret, Z.~Basrak, N.~Bastid,
  A.~Bendarag, G.~Berek, R.~\ifmmode~\check{C}\else \v{C}\fi{}aplar,
  P.~Crochet, A.~Devismes, P.~Dupieux, M.~D\ifmmode~\check{z}\else
  \v{z}\fi{}elalija, C.~Finck, Z.~Fodor, A.~Gobbi, Y.~Grishkin, O.~N. Hartmann,
  N.~Herrmann, K.~D. Hildenbrand, B.~Hong, J.~Kecskemeti, Y.~J. Kim,
  M.~Kirejczyk, P.~Koczon, M.~Korolija, R.~Kotte, T.~Kress, R.~Kutsche,
  A.~Lebedev, Y.~Leifels, W.~Neubert, D.~Pelte, M.~Petrovici, F.~Rami,
  B.~de~Schauenburg, D.~Sch\"ull, Z.~Seres, B.~Sikora, K.~S. Sim, V.~Simion,
  K.~Siwek-Wilczy\ifmmode~\acute{n}\else \'{n}\fi{}ska, V.~Smolyankin, M.~R.
  Stockmeier, G.~Stoicea, P.~Wagner, K.~Wi\ifmmode~\acute{s}\else
  \'{s}\fi{}niewski, D.~Wohlfarth, I.~Yushmanov, A.~Zhilin,
  \href{https://link.aps.org/doi/10.1103/PhysRevC.64.041604}{{D}ifferential
  directed flow in {A}u+{A}u collisions}, {P}hys. {R}ev. {C} 64 (2001) 041604.
\newblock \href {https://doi.org/10.1103/PhysRevC.64.041604}
  {\path{doi:10.1103/PhysRevC.64.041604}}.
\newline\urlprefix\url{https://link.aps.org/doi/10.1103/PhysRevC.64.041604}

\bibitem{Filippo2017}
{De Filippo, E.}, {Russotto, P.}, {Acosta, L.}, {Adamczyk, M.}, {Al-Ajlan, A.},
  {Al-Garawi, M.}, {Al-Homaidhi, S.}, {Amorini, F.}, {Auditore, L.}, {Aumann,
  T.}, {Ayyad, Y.}, {Basrak, Z.}, {Benlliure, J.}, {Boisjoli, M.}, {Boretzky,
  K.}, {Brzychczyk, J.}, {Budzanowski, A.}, {Caesar, C.}, {Cardella, G.},
  {Cammarata, P.}, {Chajecki, Z.}, {Chartier, M.}, {Chbihi, A.}, {Colonna, M.},
  {Cozma, M.D.}, {Czech, B.}, {Di Toro, M.}, {Famiano, M.}, {Gannon, S.},
  {Gaspari\'{}c, I.}, {Grassi, L.}, {Guazzoni, C.}, {Guazzoni, P.}, {Heil, M.},
  {Heilborn, L.}, {Introzzi, R.}, {Isobe, T.}, {Kezzar, K.}, {Kis, M.},
  {Krasznahorkay, A.}, {Kupny, S.}, {Kurz, N.}, {La Guidara, E.}, {Lanzalone,
  G.}, {Lasko, P.}, {Le F\`evre, A.}, {Leifels, Y.}, {Lemmon, R.C.}, {Li,
  Q.F.}, {Lombardo, I.}, {Lukasik, J.}, {Lynch, W.G.}, {Marini, P.}, {Matthews,
  Z.}, {May, L.}, {Minniti, T.}, {Mostazo, M.}, {Pagano, A.}, {Pagano, E.V.},
  {Papa, M.}, {Pawlowski, P.}, {Pirrone, S.}, {Politi, G.}, {Porto, F.},
  {Reviol, W.}, {Riccio, F.}, {Rizzo, F.}, {Rosato, E.}, {Rossi, D.}, {Santoro,
  S.}, {Sarantites, D.G.}, {Simon, H.}, {Skwirczynska, I.}, {Sosin, Z.},
  {Stuhl, L.}, {Trautmann, W.}, {Trifir\`o, A.}, {Trimarchi, M.}, {Tsang,
  M.B.}, {Verde, G.}, {Veselsky, M.}, {Vigilante, M.}, {Wang, Yongjia},
  {Wieloch, A.}, {Wigg, P.}, {Winkelbauer, J.}, {Wolter, H.H.}, {Wu, P.},
  {Yennello, S.}, {Zambon, P.}, {Zetta, L.}, {Zoric, M.},
  \href{https://doi.org/10.1051/epjconf/201713709002}{{T}he symmetry energy at
  suprasaturation density and the {A}{S}{Y}-{E}{O}{S} experiment at {G}{S}{I}},
  {E}{P}{J} {W}eb {C}onf. 137 (2017) 09002.
\newblock \href {https://doi.org/10.1051/epjconf/201713709002}
  {\path{doi:10.1051/epjconf/201713709002}}.
\newline\urlprefix\url{https://doi.org/10.1051/epjconf/201713709002}

\bibitem{le2016constraining}
A.~Le~Fevre, Y.~Leifels, W.~Reisdorf, J.~Aichelin, C.~Hartnack, Constraining
  the nuclear matter equation of state around twice saturation density, Nuclear
  Physics A 945 (2016) 112--133.

\bibitem{Russotto2023}
P.~Russotto, M.~D. Cozma, E.~D. Filippo, A.~L. F{\`{e}}vre, Y.~Leifels,
  J.~{\L}ukasik, \href{https://doi.org/10.1007\%2Fs40766-023-00039-4}{Studies
  of the equation-of-state of nuclear matter by heavy-ion collisions at
  intermediate energy in the multi-messenger era}, La Rivista del Nuovo Cimento
  46~(1) (2023) 1--70.
\newblock \href {https://doi.org/10.1007/s40766-023-00039-4}
  {\path{doi:10.1007/s40766-023-00039-4}}.
\newline\urlprefix\url{https://doi.org/10.1007\%2Fs40766-023-00039-4}

\bibitem{Stoicea2004}
G.~Stoicea, M.~Petrovici, A.~Andronic, N.~Herrmann, J.~P. Alard, Z.~Basrak,
  V.~Barret, N.~Bastid, R.~\ifmmode~\check{C}\else \v{C}\fi{}aplar, P.~Crochet,
  P.~Dupieux, M.~D\ifmmode~\check{z}\else \v{z}\fi{}elalija, Z.~Fodor,
  O.~Hartmann, K.~D. Hildenbrand, B.~Hong, J.~Kecskemeti, Y.~J. Kim,
  M.~Kirejczyk, M.~Korolija, R.~Kotte, T.~Kress, A.~Lebedev, Y.~Leifels,
  X.~Lopez, M.~Merschmeier, W.~Neubert, D.~Pelte, F.~Rami, W.~Reisdorf,
  D.~Sch\"ull, Z.~Seres, B.~Sikora, K.~S. Sim, V.~Simion,
  K.~Siwek-Wilczy\ifmmode~\acute{n}\else \'{n}\fi{}ska, V.~Smolyankin,
  M.~Stockmeier, K.~Wi\ifmmode~\acute{s}\else \'{s}\fi{}niewski, D.~Wohlfarth,
  I.~Yushmanov, A.~Zhilin, P.~Danielewicz,
  \href{https://link.aps.org/doi/10.1103/PhysRevLett.92.072303}{Azimuthal
  dependence of collective expansion for symmetric heavy-ion collisions}, Phys.
  Rev. Lett. 92 (2004) 072303.
\newblock \href {https://doi.org/10.1103/PhysRevLett.92.072303}
  {\path{doi:10.1103/PhysRevLett.92.072303}}.
\newline\urlprefix\url{https://link.aps.org/doi/10.1103/PhysRevLett.92.072303}

\bibitem{Reisdorf2012}
W.~Reisdorf, Y.~Leifels, A.~Andronic, R.~Averbeck, V.~Barret, Z.~Basrak,
  N.~Bastid, M.~Benabderrahmane, R.~Čaplar, P.~Crochet, P.~Dupieux,
  M.~Dželalija, Z.~Fodor, P.~Gasik, Y.~Grishkin, O.~Hartmann, N.~Herrmann,
  K.~Hildenbrand, B.~Hong, T.~Kang, J.~Kecskemeti, Y.~Kim, M.~Kirejczyk,
  M.~Kiš, P.~Koczoń, M.~Korolija, R.~Kotte, T.~Kress, A.~Lebedev, X.~Lopez,
  T.~Matulewicz, M.~Merschmeyer, W.~Neubert, M.~Petrovici, K.~Piasecki,
  F.~Rami, M.~Ryu, A.~Schüttauf, Z.~Seres, B.~Sikora, K.~Sim, V.~Simion,
  K.~Siwek-Wilczyńska, V.~Smolyankin, M.~Stockmeier, G.~Stoicea, Z.~Tymiński,
  K.~Wiśniewski, D.~Wohlfarth, Z.~Xiao, H.~Xu, I.~Yushmanov, A.~Zhilin,
  \href{https://www.sciencedirect.com/science/article/pii/S0375947411006877}{Systematics
  of azimuthal asymmetries in heavy ion collisions in the 1a gev regime},
  Nuclear Physics A 876 (2012) 1--60.
\newblock \href
  {https://doi.org/https://doi.org/10.1016/j.nuclphysa.2011.12.006}
  {\path{doi:https://doi.org/10.1016/j.nuclphysa.2011.12.006}}.
\newline\urlprefix\url{https://www.sciencedirect.com/science/article/pii/S0375947411006877}

\bibitem{Voloshin1996}
S.~Voloshin, Y.~Zhang, \href{https://doi.org/10.1007/s002880050141}{Flow study
  in relativistic nuclear collisions by fourier expansion of azimuthal particle
  distributions}, Zeitschrift f{\"u}r Physik C Particles and Fields 70~(4)
  (1996) 665--671.
\newblock \href {https://doi.org/10.1007/s002880050141}
  {\path{doi:10.1007/s002880050141}}.
\newline\urlprefix\url{https://doi.org/10.1007/s002880050141}

\bibitem{Poskanzer1998}
A.~M. Poskanzer, S.~A. Voloshin,
  \href{https://link.aps.org/doi/10.1103/PhysRevC.58.1671}{{M}ethods for
  analyzing anisotropic flow in relativistic nuclear collisions}, {P}hys.
  {R}ev. {C} 58 (1998) 1671--1678.
\newblock \href {https://doi.org/10.1103/PhysRevC.58.1671}
  {\path{doi:10.1103/PhysRevC.58.1671}}.
\newline\urlprefix\url{https://link.aps.org/doi/10.1103/PhysRevC.58.1671}

\bibitem{ONO2019}
A.~Ono,
  \href{https://www.sciencedirect.com/science/article/pii/S0146641018300863}{{D}ynamics
  of clusters and fragments in heavy-ion collisions}, {P}rogress in {P}article
  and {N}uclear {P}hysics 105 (2019) 139--179.
\newblock \href {https://doi.org/https://doi.org/10.1016/j.ppnp.2018.11.001}
  {\path{doi:https://doi.org/10.1016/j.ppnp.2018.11.001}}.
\newline\urlprefix\url{https://www.sciencedirect.com/science/article/pii/S0146641018300863}

\bibitem{Donigus2020}
B.~D{\"o}nigus,
  \href{https://doi.org/10.1140/epja/s10050-020-00275-w}{{S}elected highlights
  of the production of light (anti-)(hyper-)nuclei in ultra-relativistic
  heavy-ion collisions}, {T}he {E}uropean {P}hysical {J}ournal {A} 56~(11)
  (2020) 280.
\newblock \href {https://doi.org/10.1140/epja/s10050-020-00275-w}
  {\path{doi:10.1140/epja/s10050-020-00275-w}}.
\newline\urlprefix\url{https://doi.org/10.1140/epja/s10050-020-00275-w}

\bibitem{Gossiaux1997}
P.~Gossiaux, R.~Puri, C.~Hartnack, J.~Aichelin,
  \href{https://www.sciencedirect.com/science/article/pii/S0375947497001759}{The
  multifragmentation of spectator matter}, Nuclear Physics A 619~(3) (1997)
  379--390.
\newblock \href {https://doi.org/https://doi.org/10.1016/S0375-9474(97)00175-9}
  {\path{doi:https://doi.org/10.1016/S0375-9474(97)00175-9}}.
\newline\urlprefix\url{https://www.sciencedirect.com/science/article/pii/S0375947497001759}

\bibitem{Chen20212}
X.~Chen, Y.~Zhang, Z.~li,
  \href{https://doi.org/10.1088/1674-1137/abfb51}{{T}heoretical uncertainties
  on the extraction of in-medium {{N}{N}} cross sections by different {P}auli
  blocking algorithms}, {C}hinese {P}hysics {C} (Apr 2021).
\newblock \href {https://doi.org/10.1088/1674-1137/abfb51}
  {\path{doi:10.1088/1674-1137/abfb51}}.
\newline\urlprefix\url{https://doi.org/10.1088/1674-1137/abfb51}

\bibitem{LI2022}
P.~Li, Y.~Wang, Q.~Li, H.~Zhang,
  \href{https://www.sciencedirect.com/science/article/pii/S0370269322001538}{Accessing
  the in-medium effects on nucleon-nucleon elastic cross section with
  collective flows and nuclear stopping}, Physics Letters B 828 (2022) 137019.
\newblock \href
  {https://doi.org/https://doi.org/10.1016/j.physletb.2022.137019}
  {\path{doi:https://doi.org/10.1016/j.physletb.2022.137019}}.
\newline\urlprefix\url{https://www.sciencedirect.com/science/article/pii/S0370269322001538}

\bibitem{REISDORF2010}
W.~Reisdorf, A.~Andronic, R.~Averbeck, M.~Benabderrahmane, O.~Hartmann,
  N.~Herrmann, K.~Hildenbrand, T.~Kang, Y.~Kim, M.~Kiš, P.~Koczoń, T.~Kress,
  Y.~Leifels, M.~Merschmeyer, K.~Piasecki, A.~Schüttauf, M.~Stockmeier,
  V.~Barret, Z.~Basrak, N.~Bastid, R.~Čaplar, P.~Crochet, P.~Dupieux,
  M.~Dželalija, Z.~Fodor, P.~Gasik, Y.~Grishkin, B.~Hong, J.~Kecskemeti,
  M.~Kirejczyk, M.~Korolija, R.~Kotte, A.~Lebedev, X.~Lopez, T.~Matulewicz,
  W.~Neubert, M.~Petrovici, F.~Rami, M.~Ryu, Z.~Seres, B.~Sikora, K.~Sim,
  V.~Simion, K.~Siwek-Wilczyńska, V.~Smolyankin, G.~Stoicea, Z.~Tymiński,
  K.~Wiśniewski, D.~Wohlfarth, Z.~Xiao, H.~Xu, I.~Yushmanov, A.~Zhilin,
  \href{https://www.sciencedirect.com/science/article/pii/S0375947410006810}{{S}ystematics
  of central heavy ion collisions in the 1{A} {G}e{V} regime}, {N}uclear
  {P}hysics {A} 848~(3) (2010) 366--427.
\newblock \href
  {https://doi.org/https://doi.org/10.1016/j.nuclphysa.2010.09.008}
  {\path{doi:https://doi.org/10.1016/j.nuclphysa.2010.09.008}}.
\newline\urlprefix\url{https://www.sciencedirect.com/science/article/pii/S0375947410006810}

\bibitem{GAITANOS2005}
T.~Gaitanos, C.~Fuchs, H.~Wolter,
  \href{https://www.sciencedirect.com/science/article/pii/S0370269305001486}{{N}uclear
  stopping and flow in heavy-ion collisions and the in-medium {N}{N} cross
  section}, {P}hysics {L}etters {B} 609~(3) (2005) 241--246.
\newblock \href
  {https://doi.org/https://doi.org/10.1016/j.physletb.2005.01.069}
  {\path{doi:https://doi.org/10.1016/j.physletb.2005.01.069}}.
\newline\urlprefix\url{https://www.sciencedirect.com/science/article/pii/S0370269305001486}

\bibitem{Barney2021}
J.~Barney, J.~Estee, W.~G. Lynch, T.~Isobe, G.~Jhang, M.~Kurata-Nishimura,
  A.~B. McIntosh, T.~Murakami, R.~Shane, S.~Tangwancharoen, M.~B. Tsang,
  G.~Cerizza, M.~Kaneko, J.~W. Lee, C.~Y. Tsang, R.~Wang, C.~Anderson, H.~Baba,
  Z.~Chajecki, M.~Famiano, R.~Hodges-Showalter, B.~Hong, T.~Kobayashi,
  P.~Lasko, J.~Łukasik, N.~Nakatsuka, R.~Olsen, H.~Otsu, P.~Pawłowski,
  K.~Pelczar, H.~Sakurai, C.~Santamaria, H.~Setiawan, A.~Taketani, J.~R.
  Winkelbauer, Z.~Xiao, S.~J. Yennello, J.~Yurkon, Y.~Zhang,
  \href{https://www.osti.gov/biblio/1797552}{{T}he {S}$\pi${R}{I}{T} time
  projection chamber}, Review of Scientific Instruments 92~(6) (6 2021).
\newblock \href {https://doi.org/10.1063/5.0041191}
  {\path{doi:10.1063/5.0041191}}.
\newline\urlprefix\url{https://www.osti.gov/biblio/1797552}

\bibitem{Kaneko2021}
M.~Kaneko, T.~Murakami, T.~Isobe, M.~Kurata-Nishimura, A.~Ono, N.~Ikeno,
  J.~Barney, G.~Cerizza, J.~Estee, G.~Jhang, J.~Lee, W.~Lynch, C.~Santamaria,
  C.~Tsang, M.~Tsang, R.~Wang, D.~Ahn, L.~Atar, T.~Aumann, H.~Baba,
  K.~Boretzky, J.~Brzychczyk, N.~Chiga, N.~Fukuda, I.~Gašparić, B.~Hong,
  A.~Horvat, T.~Ichihara, K.~Ieki, N.~Inabe, Y.~Kim, T.~Kobayashi, Y.~Kondo,
  P.~Lasko, H.~Lee, Y.~Leifels, J.~Łukasik, J.~Manfredi, A.~McIntosh,
  P.~Morfouace, T.~Nakamura, N.~Nakatsuka, S.~Nishimura, R.~Olsen, H.~Otsu,
  P.~Pawłowski, K.~Pelczar, D.~Rossi, H.~Sakurai, H.~Sato, H.~Scheit,
  R.~Shane, Y.~Shimizu, H.~Simon, T.~Sumikama, D.~Suzuki, H.~Suzuki, H.~Takeda,
  S.~Tangwancharoen, Y.~Togano, H.~Törnqvist, Z.~Xiao, S.~Yennello, J.~Yurkon,
  Y.~Zhang,
  \href{https://www.sciencedirect.com/science/article/pii/S0370269321006213}{{R}apidity
  distributions of {Z} = 1 isotopes and the nuclear symmetry energy from
  {S}n+{S}n collisions with radioactive beams at 270 {M}e{V}/nucleon},
  {P}hysics {L}etters {B} 822 (2021) 136681.
\newblock \href
  {https://doi.org/https://doi.org/10.1016/j.physletb.2021.136681}
  {\path{doi:https://doi.org/10.1016/j.physletb.2021.136681}}.
\newline\urlprefix\url{https://www.sciencedirect.com/science/article/pii/S0370269321006213}

\bibitem{ZHANG2014}
Y.~Zhang, M.~Tsang, Z.~Li, H.~Liu,
  \href{https://www.sciencedirect.com/science/article/pii/S0370269314001865}{{C}onstraints
  on nucleon effective mass splitting with heavy ion collisions}, {P}hys.
  {L}ett. {B} 732 (2014) 186--190.
\newblock \href
  {https://doi.org/https://doi.org/10.1016/j.physletb.2014.03.030}
  {\path{doi:https://doi.org/10.1016/j.physletb.2014.03.030}}.
\newline\urlprefix\url{https://www.sciencedirect.com/science/article/pii/S0370269314001865}

\bibitem{ZHANG2015}
Y.~Zhang, M.~Tsang, Z.~Li,
  \href{https://www.sciencedirect.com/science/article/pii/S0370269315005742}{Covariance
  analysis of symmetry energy observables from heavy ion collision}, Physics
  Letters B 749 (2015) 262--266.
\newblock \href
  {https://doi.org/https://doi.org/10.1016/j.physletb.2015.07.064}
  {\path{doi:https://doi.org/10.1016/j.physletb.2015.07.064}}.
\newline\urlprefix\url{https://www.sciencedirect.com/science/article/pii/S0370269315005742}

\bibitem{Skyrme1956}
T.~H.~R. Skyrme, \href{https://doi.org/10.1080/14786435608238186}{Cvii. the
  nuclear surface}, The Philosophical Magazine: A Journal of Theoretical
  Experimental and Applied Physics 1~(11) (1956) 1043--1054.
\newblock \href
  {http://arxiv.org/abs/https://doi.org/10.1080/14786435608238186}
  {\path{arXiv:https://doi.org/10.1080/14786435608238186}}, \href
  {https://doi.org/10.1080/14786435608238186}
  {\path{doi:10.1080/14786435608238186}}.
\newline\urlprefix\url{https://doi.org/10.1080/14786435608238186}

\bibitem{Vautherin1972}
D.~Vautherin, D.~M. Brink,
  \href{https://link.aps.org/doi/10.1103/PhysRevC.5.626}{Hartree-fock
  calculations with skyrme's interaction. i. spherical nuclei}, Phys. Rev. C 5
  (1972) 626--647.
\newblock \href {https://doi.org/10.1103/PhysRevC.5.626}
  {\path{doi:10.1103/PhysRevC.5.626}}.
\newline\urlprefix\url{https://link.aps.org/doi/10.1103/PhysRevC.5.626}

\bibitem{Chabanat1997}
E.~Chabanat, P.~Bonche, P.~Haensel, J.~Meyer, R.~Schaeffer,
  \href{https://www.sciencedirect.com/science/article/pii/S0375947497005964}{A
  skyrme parametrization from subnuclear to neutron star densities}, Nuclear
  Physics A 627~(4) (1997) 710--746.
\newblock \href {https://doi.org/https://doi.org/10.1016/S0375-9474(97)00596-4}
  {\path{doi:https://doi.org/10.1016/S0375-9474(97)00596-4}}.
\newline\urlprefix\url{https://www.sciencedirect.com/science/article/pii/S0375947497005964}

\bibitem{Agrawal2005}
B.~K. Agrawal, S.~Shlomo, V.~K. Au,
  \href{https://link.aps.org/doi/10.1103/PhysRevC.72.014310}{Determination of
  the parameters of a skyrme type effective interaction using the simulated
  annealing approach}, Phys. Rev. C 72 (2005) 014310.
\newblock \href {https://doi.org/10.1103/PhysRevC.72.014310}
  {\path{doi:10.1103/PhysRevC.72.014310}}.
\newline\urlprefix\url{https://link.aps.org/doi/10.1103/PhysRevC.72.014310}

\bibitem{Chen2009}
L.-W. Chen, B.-J. Cai, C.~M. Ko, B.-A. Li, C.~Shen, J.~Xu,
  \href{https://link.aps.org/doi/10.1103/PhysRevC.80.014322}{Higher-order
  effects on the incompressibility of isospin asymmetric nuclear matter}, Phys.
  Rev. C 80 (2009) 014322.
\newblock \href {https://doi.org/10.1103/PhysRevC.80.014322}
  {\path{doi:10.1103/PhysRevC.80.014322}}.
\newline\urlprefix\url{https://link.aps.org/doi/10.1103/PhysRevC.80.014322}

\bibitem{Cugnon1996}
J.~Cugnon, D.~L'Hôte, J.~Vandermeulen,
  \href{https://www.scopus.com/inward/record.uri?eid=2-s2.0-0030143153&doi=10.1016%2f0168-583X%2895%2901384-9&partnerID=40&md5=e6854ad674cc549da0dbc3c211cefbf0}{{S}imple
  parametrization of cross-sections for nuclear transport studies up to the
  {G}e{V} range}, {N}uclear {I}nstruments and {M}ethods in {P}hysics
  {R}esearch, {S}ection {B}: {B}eam {I}nteractions with {M}aterials and {A}toms
  111~(3-4) (1996) 215--220, cited By 95.
\newblock \href {https://doi.org/10.1016/0168-583X(95)01384-9}
  {\path{doi:10.1016/0168-583X(95)01384-9}}.
\newline\urlprefix\url{https://www.scopus.com/inward/record.uri?eid=2-s2.0-0030143153&doi=10.1016%2f0168-583X%2895%2901384-9&partnerID=40&md5=e6854ad674cc549da0dbc3c211cefbf0}

\bibitem{Patil2010}
A.~Patil, D.~Huard, C.~J. Fonnesbeck,
  \href{https://pubmed.ncbi.nlm.nih.gov/21603108}{{P}y{M}{C}: {B}ayesian
  {S}tochastic {M}odelling in {P}ython}, {J}ournal of statistical software
  35~(4) (2010) 1--81, 21603108[pmid].
\newline\urlprefix\url{https://pubmed.ncbi.nlm.nih.gov/21603108}

\bibitem{Rasmussen2005}
C.~E. Rasmussen, C.~K.~I. Williams, {G}aussian {P}rocesses for {M}achine
  {L}earning ({A}daptive {C}omputation and {M}achine {L}earning) (2005).

\bibitem{Li2022_2}
B.-A. Li, B.-J. Cai, L.-W. Chen, W.-J. Xie, J.~Xu, N.-B. Zhang, A theoretical
  overview of isospin and eos effects in heavy-ion reactions at intermediate
  energies, Il Nuovo Cimento C 45~(3) (6 2022).
\newblock \href {https://doi.org/10.1393/ncc/i2022-22054-3}
  {\path{doi:10.1393/ncc/i2022-22054-3}}.

\bibitem{Hermann2021}
H.~Wolter, M.~Colonna, D.~Cozma, P.~Danielewicz, C.~M. Ko, R.~Kumar, A.~Ono,
  M.~B. Tsang, J.~Xu, Y.-X. Zhang, E.~Bratkovskaya, Z.-Q. Feng, T.~Gaitanos,
  A.~L. Fèvre, N.~Ikeno, Y.~Kim, S.~Mallik, P.~Napolitani, D.~Oliinychenko,
  T.~Ogawa, M.~Papa, J.~Su, R.~Wang, Y.-J. Wang, J.~Weil, F.-S. Zhang, G.-Q.
  Zhang, Z.~Zhang, J.~Aichelin, W.~Cassing, L.-W. Chen, H.-G. Cheng, H.~Elfner,
  K.~Gallmeister, C.~Hartnack, S.~Hashimoto, S.~Jeon, K.~Kim, M.~Kim, B.-A. Li,
  C.-H. Lee, Q.-F. Li, Z.-X. Li, U.~Mosel, Y.~Nara, K.~Niita, A.~Ohnishi,
  T.~Sato, T.~Song, A.~Sorensen, N.~Wang, W.-J. Xie, Transport model comparison
  studies of intermediate-energy heavy-ion collisions (2022).
\newblock \href {http://arxiv.org/abs/2202.06672} {\path{arXiv:2202.06672}}.

\end{thebibliography}
\end{document}